\documentclass[reprint,groupedaddress,showpacs,preprintnumbers,nofootinbib,nobibnotes,amsmath,amssymb,aps,prd,floatfix,superscriptaddress]{revtex4-1}

\usepackage[colorlinks=false]{hyperref}
\usepackage{dcolumn}
\usepackage{bm}
\usepackage[normalem]{ulem}

\usepackage{graphicx,color,xcolor}
\usepackage[lofdepth,lotdepth]{subfig}
\usepackage{siunitx}
\usepackage{enumitem}
\usepackage[utf8]{inputenc}
\usepackage{mathrsfs}

\DeclareSIUnit{\mev}{\mega\electronvolt}

\newcommand{\nuslash}{k_\nu\!\!\!\!\!/}
\newcommand{\eslash}{k_e\!\!\!\!\!/}

\newcommand{\peslash}{K_e\!\!\!\!\!/}
\newcommand{\pnuslash}{K_\nu\!\!\!\!\!/}
\newcommand{\ls}{\lambda\sigma}
\newcommand{\Tr}{\mathrm{Tr}}
\newcommand{\ep}{\epsilon_k^p}
\newcommand{\en}{\epsilon_{k+q}^n}
\newcommand{\enk}{\epsilon_{k}^n}
\def\hdf{\textsc{HDF5}}

\def\mnras{Mon. Notices Royal Astron. Soc.} 	

\begin{document}

\title{Improved neutrino-nucleon interactions in dense and hot matter
  for numerical simulations}

\author{Micaela Oertel}\email{micaela.oertel@obspm.fr}
\affiliation{LUTH, Observatoire de Paris, Universit\'e PSL, CNRS,
  Universit\'e de Paris, 92190 Meudon, France}%
\author{Aur\'elien Pascal}\email{aurelien.pascal@obspm.fr}
\affiliation{LUTH, Observatoire de Paris, Universit\'e PSL, CNRS,
  Universit\'e de Paris, 92190 Meudon, France}%
\author{Marco Mancini}\email{marco.mancini@obspm.fr}
\affiliation{LUTH, Observatoire de Paris, Universit\'e PSL, CNRS,
  Universit\'e de Paris, 92190 Meudon, France}%
\author{J\'er\^ome Novak}\email{jerome.novak@obspm.fr}
\affiliation{LUTH, Observatoire de Paris, Universit\'e PSL, CNRS,
  Universit\'e de Paris, 92190 Meudon, France}%

\date{\today}

\begin{abstract}
  Neutrinos play an important role in compact star astrophysics:
  neutrino-heating is one of the main ingredients in core-collapse
  supernovae, neutrino-matter interactions determine the composition
  of matter in binary neutron star mergers and have among others a
  strong impact on conditions for heavy element nucleosynthesis and
  neutron star cooling is dominated by neutrino emission except for
  very old stars. Many works in the last decades have shown that in
  dense matter medium effects considerably change the neutrino-matter
  interaction rates, whereas many astrophysical simulations use
  analytic approximations which are often far from reproducing more
  complete calculations. In this work we present a scheme which allows
  to incorporate improved rates for charged current interactions,
  into simulations and show as an example some results for
  core-collapse supernovae, where a noticeable difference is found in
  the location of the neutrinospheres of the low-energy neutrinos in
  the early post-bounce phase.
\end{abstract}


\maketitle

\section{Introduction}\label{s:intro}

The first detection of gravitational waves (GWs) from a binary neutron
star (BNS) merger by the LIGO-Virgo collaboration in August 2017, the
event GW170817, in coincidence with the observation of a gamma ray
burst (GRB170817a) and an electro-magnetic counterpart established the
beginning of multi-messenger astronomy \cite{abbott_17}. Many
additional detections are expected including several BNS mergers,
during current and forthcoming campaigns by large interferometric
gravitational wave detectors. This rapidly evolving new astronomy is
revolutionizing the exploration of the universe by addressing
fundamental questions such as the nature of gravity, of dark matter,
the origin of elements heavier than iron and of properties of dense
matter in compact stars. A complete understanding of these exciting
observations will be achieved once they can be successfully confronted
to the predictions of theoretical modeling for which still many
questions remain open, among others concerning neutrino
interactions. The latter play an essential role for astrophysics of
compact objects:
\begin{enumerate}
\item The dynamics of BNS mergers only marginally depend on neutrino
  interactions. However, ejecta composition and nucleosynthesis
  conditions are very sensitive to the neutrino treatment and neutrino
  interactions.
\item The heating by neutrinos of the stalled shock wave represents a
  crucial element for the dynamics of core-collapse supernovae (CCSN),
  contributing to the explosion mechanism
\item (Proto)-neutron star cooling is dominated by neutrino emission
  for millions of years.
\end{enumerate}
Simulations of these processes are very expensive, so that currently
mostly analytic expressions for the relevant reaction rates
\cite{Bruenn:1985en, Rosswog:2003rv, Burrows:2004vq, Schmitt:2017efp}
are applied, which are, however, often based on very crude
approximations. Several corrections have been added to the original
expressions~\cite{Bruenn:1985en}, such as weak magnetism and recoil
\cite{Horowitz:2001xf}, nuclear structure corrections
\cite{Horowitz:1996ci, Bruenn:1997jv}, effective masses and chemical
potentials for nucleons in dense matter \cite{Reddy:1997yr,Roberts:2012um,
  MartinezPinedo:2012rb}, additional reactions \cite{Hannestad:1997gc,
  Fischer:2018kdt} and superfluidity in cooling neutron stars older
than several minutes \cite{Yakovlev:2000jp}. Several authors have,
however, pointed out since decades that in dense matter different
effects can additionally modify the neutrino matter interaction rates
and neutrino emissivities by orders of magnitude, in particular
nuclear correlations \cite{Burrows:1998cg, Burrows:1998ek,
  Reddy:1998hb, Navarro:1999jn, Margueron:2004sr, Horowitz:2006pj,
  Horowitz:2016gul}. (Special) relativistic effects can play an
important role, too
\cite{Leinson:2001ei,Leinson:2002bw,Roberts:2016mwj}. Most of the
above discussed modifications have been cast into correction factors
to the original analytic expressions for practical use in
simulations. The \textsc{NuLib} library by Evan
O'Connor~\cite{OConnor:2014sgn} provides the corresponding neutrino
opacities.

It is, however, not possible to provide analytic expressions taking
into account all known corrections. In view of the computational
effort, only few simulations go indeed beyond the analytical
expressions. In particular, the full phase space has been considered
in several PNS cooling simulations~\cite{Pons:1998mm,Roberts:2012um}
as well as for charged-current opacities in the spherically symmetric
CCSN simulations of \textcite{Fischer:2018kdt} and the Garching group
has implemented additionally nuclear correlations following the
simplified formalism of \textcite{Burrows:1998cg,Burrows:1998ek}, see
\textcite{Buras:2005tb,Huedepohl:2009wh}. However, as for
neutrino-nucleus reactions, where already for a long time tabulated
and accurate data on selected nuclei from microscopic calculations
accounting for nuclear correlations and thermal effects
exist~\cite{FFN_1982b, LMP_ADNDT_2001, Langanke2002, Oda1994,
  Pruet2003,Juodagalvis_NPA_2010}, it is desirable that improved rates
for neutrino-nucleon reactions become more generally available. In
\textcite{Roberts:2016mwj} a first step has been performed in this
direction, neutrino opacities for charged current reactions with the
full relativistic phase space including mean field corrections are
provided via the \textsc{nuOpac} library. In the present work we
perform a step further towards a complete neutrino toolkit, allowing
to provide state-of-the-art neutrino-matter interaction rates directly
applicable to simulations. As \textcite{Roberts:2016mwj}, we will
concentrate on charged current neutrino-nucleon interactions,
i.e. neutrino absorption and creation. We will stick to
non-relativistic kinematics, but in addition to the full phase space,
which is important at high
densities~\cite{Roberts:2016mwj,Fischer:2018kdt}, we will include
nuclear correlations via the so-called ``Random Phase
Approximation''(RPA). The impact of RPA correlations on neutrino
opacities has been studied in several works, see
e.g. Refs.~\cite{Reddy:1998hb, Burrows:1998ek}, but mostly only grey,
i.e. neutrino energy independent, correction factors to the above
cited analytic approximation have yet been implemented. As discussed
in Section~\ref{ss:results}, the importance of RPA correlations is,
however, energy dependent and a density and temperature dependent
shift in reaction thresholds is induced. It is thus obvious that the
full physics cannot be included into a grey factor and we will for the
first time provide opacities with the full dependence on neutrino
energy $E_\nu$, baryon number density $n_B$, temperature $T$ and
electron fraction $Y_e$.

We will consider thermodynamic conditions relevant for BNS mergers,
CCSN and (proto)-neutron star cooling during the first minutes which
are rather similar: hot and dense nuclear matter with different
asymmetries, i.e. proton to neutron ratios. In the central and hot
parts, matter is homogeneous, whereas in the outer regions, containing
more dilute and cold matter, nuclear clusters coexist with free
nucleons. Charged current neutrino-nucleon reactions thereby, together
with neutrino-nucleon scattering, not only control neutrino diffusion
and emission in the central part, but strongly influence the physics
close to the neutrinospheres, too, which is important for the dynamics
and the characteristics of the emerging neutrinos. Charged current
reactions on nucleons contribute critically to the heating of matter
behind the shock in a CCSN, too.

Since the aim of our work is to provide results for the interaction
rates going beyond the analytic approximations, complete opacity data
as function of $E_\nu, n_B, T$ and $Y_e$ as calculated within the
present work can be found in tabular form on the \textsc{Compose} data
base~\cite{Typel:2013rza}\footnote{\url{https://compose.obspm.fr}}
together with the underlying equation of state (EoS) data. It has been
emphasized many times that it is important to determine
neutrino-nucleon interactions coherently with the underlying EoS,
i.e. employing the same model for nuclear interactions, see
e.g.~\cite{Reddy:1998hb}. In contrast to \textcite{Roberts:2016mwj}
who provide routines and \textcite{Fischer:2018kdt} who use a phase
space integration ``on the fly'' during spherically symmetric
simulations, we have chosen to directly provide opacity tables since
the numerical calculations to obtain the rates in RPA (and probably
any other scheme taking into account nuclear correlations) are much
more time consuming than the simpler phase space integrations due to
the existence of collective excitations in the nuclear response, see
e.g. \cite{Hernandez:1999zz}. They have thus
anyway to be tabulated before they can be implemented into
simulations. We will show that our scheme can indeed be applied to
simulations and perform CCSN simulations with fully energy dependent
RPA correlations. The impact on the early post bounce evolution is
discussed.

The paper is organized as follows. In Section~\ref{s:opacities},
neutrino opacities from charged current neutrino-nucleon interactions
are computed. After briefly recalling the formalism in
Sec.~\ref{ss:formalism}, we compute in Sec.~\ref{ss:polarization} the
polarization function within different approximations devised in
previous works, some of them going beyond the standard (elastic) one. Results
concerning neutrino opacities in these different approximations are
given in Sec.~\ref{ss:results}. Section~\ref{s:ccsn} shows some
outputs from CCSN simulations using these neutrino opacities
in the solution of neutrino transport. All these points are summarized
and discussed in Section~\ref{s:discussion}.

\section{Charged current neutrino opacities}\label{s:opacities}

As mentioned in the introduction, different approximations have been
considered to derive neutrino opacities from charged current
neutrino-nucleon interactions. The elastic
approximation~\cite{Bruenn:1985en} consists in neglecting any momentum
transfer to the nucleons, assuming non-interacting nucleons and
approximating the nucleonic form factors by lowest order constants. In
\textcite{Horowitz:2001xf}, several corrections are introduced and
corresponding analytic expressions are derived: (i) a momentum
dependence of the nucleonic form factors which becomes important for
energies close to the relevant scale of 1 GeV and which can thus
safely be neglected in our case, (ii) weak magnetism corrections to
the nucleonic form factors which are proportional to the difference in
proton and neutron magnetic moment. These corrections are relevant at
any density and can be of the order of 10\%, (iii) phase space
corrections up to order $E_\nu/m_i$, where $m_i$ are free nucleon masses
and $E_\nu$ the neutrino energy.

The seminal work of Refs.~\cite{Reddy:1997yr, Reddy:1998hb,
  Burrows:1998ek} in the late 1990's discusses the effect of nuclear
interactions on the opacities as well in mean field approximation as in
RPA. The latter is a method widely used in nuclear physics in order to
account for nuclear correlations beyond mean field. It sums up
particle-hole excitations of the nuclear medium within the long range
collective (linear) response. At low densities, mean field is
recovered. RPA correlations can reduce opacities by up to a factor
five in high density matter~\cite{Reddy:1998hb, Burrows:1998ek,
  Dzhioev:2018ovi}.

More recently, interactions in dense asymmetric matter have regained
interest since it has been pointed out that they lead to a difference
in proton and neutron single particle energies which can be of the
order of several tens of MeV and can have sizable consequences for charged
current opacities. Several authors have investigated the impact of
these interactions by introducing effective masses and chemical
potentials calculated from mean field into the elastic approximation
opacity on proto-neutron star cooling~\cite{Roberts:2012um,
  MartinezPinedo:2012rb}. Since the effect is opposite for neutrinos
and anti-neutrinos, the energy difference between neutrinos and
anti-neutrinos during a CCSN is enhanced, allowing for more neutron
rich ejecta in CCSN neutrino driven winds with consequences for
nucleosynthesis~\cite{Roberts:2012um,
  MartinezPinedo:2012rb}. \textcite{Roberts:2016mwj} have incorporated
these mean field effects within full relativistic kinematics and have
shown that in particular at high densities, when the transferred
momentum becomes large, opacities are altered by a factor of a few.

In the following section~\ref{ss:formalism} we will present the
general formalism before deriving explicit expressions within
different approximations and discussing numerical results.

\subsection{Formalism}\label{ss:formalism}

In this section we will derive expressions for neutrino opacities
arising from the following reactions for neutrino
\begin{equation}
  p+e^- \leftrightarrow n + \nu_e  \quad \quad p \leftrightarrow n + e^+ + \nu_e
\end{equation}
and anti-neutrino opacities
\begin{equation}
  n \leftrightarrow p + e^- + \bar{\nu}_e  \quad \quad n + e^+
  \leftrightarrow p + \bar{\nu}_e
\end{equation}
following
Refs.~\cite{Sedrakian:1999jh,Schmitt:2005wg}. Let us consider a general
process (creation/absorption) with an incoming/outgoing nucleon and an
incoming/outgoing lepton, where one of the leptons is a neutrino.
The different reaction rates can be calculated from the kinetic
equation for the neutrino Green's function $G_\nu^{>,<}$:
\begin{eqnarray}
  i \partial^\lambda_X \Tr[\gamma_\lambda G^{<}_\nu(X,k_{\nu})] =
  &-&\Tr [G^{>}_\nu(X,k_{\nu}) \Sigma^{<} (X,k_{\nu}) \nonumber \\
  && - \Sigma^{>}(X,k_{\nu}) G_\nu^{<}(X,k_{\nu})]~,\label{eq:greens}
\end{eqnarray}
with $k_{\nu}$ the neutrino four-momentum and assuming that the
neutrino Green's function is a slowly varying function of the
space-time coordinate $X = (t,\vec{x})$. Neutrino self energies are
denoted by $\Sigma^{>,<}$ and $\gamma^\mu$ are the standard gamma
matrices. Close to equilibrium and for a spatially homogeneous system,
which is the case for our problem, we can write for the Green's
function
\begin{eqnarray}
i \,G_\nu^{<} &=& - (\nuslash + \mu_\nu \gamma^0) \frac{\pi}{E_\nu}
                  \{f_\nu(t,\vec{k_{\nu}}) \delta(k_{\nu}^0 + \mu_\nu
                  - E_\nu)\nonumber \\
&& - (1 - f_{\bar{\nu}}(t,-\vec{k_{\nu}})) \delta(k_{\nu}^0 + \mu_\nu + E_\nu)\} \nonumber \\
i \,G_\nu^{>} &=& (\nuslash + \mu_\nu \gamma^0) \frac{\pi}{E_\nu} \{(1
                  - f_\nu(t,\vec{k_{\nu}})) \delta(k_{\nu}^0 + \mu_\nu
                  - E_\nu) \nonumber \\
&& - f_{\bar{\nu}}(t,-\vec{k_{\nu}}) \delta(k_{\nu}^0 + \mu_\nu +
   E_\nu)\} ~, \label{eq:nupropagator}
\end{eqnarray}
where $\mu_\nu$ denotes the (equilibrium) neutrino chemical potential,
$E_\nu$ the (on-shell) neutrino energy and $f_{\nu,\bar{\nu}}(t,\vec{k_{\nu}})$ the
(anti-)neutrino distribution functions. A slashed four-momentum,
e.g. $\nuslash$, indicates the contraction of the corresponding
four-momentum with the gamma matrices.

The neutrino self energies are calculated in lowest order as follows
\begin{eqnarray}
  \Sigma^{<}(t,k_{\nu}) &=& \frac{G_F^2 V_{ud}^2}{2} \int \frac{d^4 k_e}{( 2\pi)^4}  \gamma^\lambda (1- \gamma_5)\nonumber \\ &\times&  (-i G^{<}(k_e)) \gamma^\sigma (1-\gamma_5) \Pi^{>}_{\lambda\sigma}(k_e-k_{\nu})~,
  \label{eq:nuself}
\end{eqnarray}
and analogously for $\Sigma^{>}$. $G_F$ denotes here the Fermi
coupling constant and $V_{ud}$ the quark mixing matrix element
entering the charged current processes with nucleons. $G^{<}(k_e)$
stands for the electron/positron Green's function with momentum
$k_e$ and the polarization functions $\Pi^{>,<}$ are the $W$-boson
self-energies which in the present context with energies maximally of
the order hundreds of MeV can be safely evaluated from Fermi theory.

For better readability, we will focus the following derivations on
electronic reactions and only give the full final expressions for
positronic processes. Combining Eqs.~(\ref{eq:nupropagator},
\ref{eq:nuself}), and inserting the explicit expression for the
electron Green's function, the traces on the right hand side of
Eq.~(\ref{eq:greens}) become
\begin{align}
  \Tr[&G^{>}_\nu\Sigma^{<}] = - i\frac{G_F^2 V_{ud}^2}{2} \int
        \frac{d^4 k_e} {(2 \pi)^4}\frac{\pi^2}{E_\nu E_e}
        \Pi^{(>)}_{\lambda\sigma} \times \nonumber \\
      & \Tr[ (\nuslash + \mu_\nu \gamma_0)\gamma^\lambda (1-\gamma_5)
        (\eslash +m_e + \mu_e \gamma_0) \gamma^\sigma (1- \gamma_5)] \nonumber \\
      & \times f_{e} \delta(k_e^0 + \mu_e - E_e) \{(1 -f_\nu)\delta
        (k_{\nu}^0 + \mu_\nu - E_\nu) \nonumber \\
      & \qquad \qquad \qquad \qquad \qquad{} - f_{\bar\nu}
        \delta(k_{\nu}^0 + \mu_\nu + E_\nu) \}
\end{align}
and analogously for $\Tr[G^{<} \Sigma^{>}]$ ($m_e$ and $\mu_e$ are
electron mass and chemical potential, respectively).
The trace on the left hand side of Eq.~(\ref{eq:greens}) can be evaluated as
\begin{align}
i \Tr[\gamma^0 G_\nu^{<}] = - 4 (k_{\nu}^0 + \mu_\nu)\frac{\pi}{E_\nu}
  \{f_\nu &\delta(k_{\nu}^0 + \mu_\nu - E_\nu) \nonumber \\ -
  (1-f_{\bar{\nu}}) &\delta(k_{\nu}^0 + \mu_\nu + E_\nu)\}~.
\end{align}
After integration over the zero-component of the neutrino momentum, we
get the following expression for the time derivative of the
(anti-)neutrino distribution function:
\begin{align}
\frac{\partial }{\partial t} f_\nu =& -i \frac{G_F^2 V_{ud}^2}{16} \int \frac{d^3 k_e}{ (2 \pi)^3} \frac{1}{E_e E_\nu} L^{\lambda\sigma}\times \nonumber \\ &\{ (1- f_\nu) f_e \Pi^{>}_{\lambda\sigma}(q)
- f_\nu (1- f_e) \Pi^{<}_{\lambda\sigma}(q)\}~,\nonumber \\
\frac{\partial }{\partial t} f_{\bar\nu} =& -i \frac{G_F^2 V_{ud}^2}{16} \int \frac{d^3 k_e}{ (2 \pi)^3} \frac{1}{E_e E_\nu} L^{\lambda\sigma} \times \nonumber \\ &\{ (1- f_{\bar{\nu}}) (1-f_e) \Pi^{<}_{\lambda\sigma}(\bar{q})
- f_{\bar\nu} f_e \Pi^{>}_{\lambda\sigma}(\bar{q})\}~,
\label{eq:nuderiv}
\end{align}
where $q = (E_e - E_\nu - \mu_e + \mu_\nu, \vec{k_e} - \vec{k_{\nu}}), \bar{q}= (E_e + E_\nu - \mu_e + \mu_\nu,\vec{k_e} + \vec{k_{\nu}})$. The lepton tensor $L^{\ls}$ only depends on electron and neutrino energies and momenta, not on the chemical potentials:
\begin{equation}
L^{\lambda\sigma} = \Tr[ (\peslash + m_e) \gamma^\sigma (1- \gamma_5) \pnuslash\ \gamma^\lambda (1-\gamma_5)]~,
\label{eq:ltensor}
\end{equation}
where $K_e = (E_e,\vec{k_e}), K_\nu = (E_\nu,\vec{k_{\nu}})$ with
on-shell energies. The forward and backward polarization functions can
be related to the retarded one in the following way
\begin{eqnarray}
\Pi_{\lambda\sigma}^{>}(q) &=& - 2 i (1 + f_B (q_0)) \mathrm{Im} \Pi_{\lambda\sigma}^R(q) \nonumber \\
\Pi_{\lambda\sigma}^{<}(q) &=& - 2 i f_B (q_0) \mathrm{Im} \Pi_{\lambda\sigma}^R(q)~,
\label{eq:retarded}
\end{eqnarray}
where $f_B$ denotes the Bose-Einstein distribution function, $f_B(q_0)
= 1/(e^{q_0/T} -1)$. The electron distribution function is described
by a Fermi-Dirac distribution, $f_e = f_F(E_e) = 1/(e^{(E_e -
  \mu_e)/T} + 1)$. Inserting Eq.~(\ref{eq:retarded}) in
Eq.~(\ref{eq:nuderiv}), the change in the (anti-)neutrino distribution
function due to electronic processes can finally be written as
\begin{align}
\frac{\partial }{\partial t} f_\nu =
& - \frac{G_F^2 V_{ud}^2}{8} \int \frac{d^3 k_e}{ (2 \pi)^3}
  \frac{1}{E_e E_\nu} L^{\lambda\sigma} \mathrm{Im}
  \Pi^{R}_{\lambda\sigma}(q) \times \nonumber \\
&\{ (1-f_\nu) f_F(E_e - \mu_e) (1 + f_B(q_0)) \nonumber \\
& \qquad{} - f_\nu (1- f_F (E_e - \mu_e)) f_B(q_0)\} ~, \nonumber \\
\frac{\partial }{\partial t} f_{\bar\nu} =
& - \frac{G_F^2 V_{ud}^2}{8} \int \frac{d^3 k_e}{ (2\pi)^3}
  \frac{1}{E_e E_\nu} L^{\lambda\sigma} \mathrm{Im}
  \Pi^{R}_{\lambda\sigma}(\bar{q})\times \nonumber \\
&\{ (1-f_{\bar\nu})(1-f_F(E_e - \mu_e)) f_B (\bar{q_0}) \nonumber \\
& \qquad{} - f_{\bar\nu} f_F (E_e - \mu_e) (1+f_B(\bar{q_0}))\}~. \label{eq:dfnudt}
\end{align}
From this change in (anti-)neutrino distribution function we can
deduce the (anti-)neutrino emissivity $j$ and the inverse mean free
path $1/\lambda$ which are related to the creation and absorption
rates via
\begin{align}
\frac{\partial }{\partial t} f_\nu &= j (E_\nu) \,(1- f_\nu) -
                                     \frac{1}{\lambda(E_\nu)} \,f_\nu
                                     \nonumber \\
\frac{\partial }{\partial t} f_{\bar{\nu}} &= \bar{\jmath} (E_\nu)
                                             \,(1- f_{\bar{\nu}}) -
                                             \frac{1}{\bar{\lambda}(E_\nu)}
                                             \,f_{\bar{\nu}}
\label{eq:jlambda}
\end{align}
as
\begin{align}
j(E_\nu) =& - \frac{G_F^2 V_{ud}^2}{8} \int \frac{d^3 k_e}{ (2 \pi)^3} \frac{1}{E_e E_\nu} L^{\lambda\sigma} \mathrm{Im} \Pi^{R}_{\lambda\sigma}(q) \times \nonumber \\ & f_F(E_e - \mu_e) (1 + f_B (q_0)) +\mathrm{positronic}\nonumber \\
\frac{1}{\lambda(E_\nu)} =& - \frac{G_F^2 V_{ud}^2}{8} \int \frac{d^3 k_e}{ (2 \pi)^3} \frac{1}{E_e E_\nu} L^{\lambda\sigma} \mathrm{Im} \Pi^{R}_{\lambda\sigma}(q) \times \nonumber \\ & (1- f_F (E_e - \mu_e)) f_B(q_0) + \mathrm{positronic}
\label{eq:nuemissivity}
\end{align}
for neutrinos and
\begin{align}
\bar{\jmath}(E_\nu) =& - \frac{G_F^2 V_{ud}^2}{8} \int \frac{d^3 k_e}{ (2 \pi)^3} \frac{1}{E_e E_\nu} L^{\lambda\sigma} \mathrm{Im} \Pi^{R}_{\lambda\sigma}(\bar{q}) \times \nonumber \\ &(1- f_F(E_e - \mu_e)) f_B (\bar{q_0}) +\mathrm{positronic}\nonumber \\
\frac{1}{\bar{\lambda}(E_\nu)} =& - \frac{G_F^2 V_{ud}^2}{8} \int \frac{d^3 k_e}{ (2 \pi)^3} \frac{1}{E_e E_\nu} L^{\lambda\sigma} \mathrm{Im} \Pi^{R}_{\lambda\sigma}(\bar{q}) \times \nonumber \\ & f_F (E_e - \mu_e) (1+ f_B(\bar{q_0})) + \mathrm{positronic}
\label{eq:nubaremissivity}
\end{align}
for anti-neutrinos. The above expressions only explicitly contain the
contribution of electronic processes and positronic ones have to be
added in order to obtain the complete emissivity and mean
free path. The latter can easily be derived, see
appendix~\ref{app:opacities}.
The properties
\begin{align}
  & f_F(E_e-\mu_e) (1+f_B(q_0)) =\nonumber \\  &\quad f_B(q_0) (1-f_F(E_e-\mu_e)) \exp((-E_\nu + \mu_\nu)/T)~,\nonumber \\
  & f_F(E_e-\mu_e) (1+f_B(\bar{q_0})) =\nonumber \\  &\quad f_B(\bar{q_0}) (1-f_F(E_e-\mu_e)) \exp((E_\nu + \mu_\nu)/T)
  \label{eq:detailedbalance}
\end{align}
relate emissivity and inverse mean free path and reflect detailed
balance. Similar properties reflect detailed balance for processes
with positrons, see appendix~\ref{app:opacities}. It is common to
introduce the absorption opacity corrected for stimulated absorption,
see e.g.~\cite{Burrows:2004vq},
\begin{align}
  \kappa_a^*(E_\nu) &= \frac{1}{1 - f_F(E_\nu-\mu_\nu)} \frac{1}{\lambda(E_\nu)} = j(E_\nu) + \frac{1}{\lambda(E_\nu)}~, \nonumber \\
  \bar{\kappa_a^*}(E_\nu) &= \frac{1}{1 - f_F(E_\nu+\mu_\nu)} \frac{1}{\bar{\lambda}(E_\nu)} = \bar{\jmath}(E_\nu) + \frac{1}{\bar{\lambda}(E_\nu)}~,\nonumber \\
\end{align}
where $f_F(E_\nu \pm \mu_\nu)$ is the equilibrium (anti-)neutrino
distribution function. Speaking about opacities below, we will always
refer to $\kappa_a^*$ and $\bar{\kappa_a^*}$ for neutrinos or
anti-neutrinos, respectively, containing the contributions from both
electronic and positronic processes.

The leptonic part in
Eqs.~(\ref{eq:nuemissivity},\ref{eq:nubaremissivity}) is evaluated
straight-forwardly, whereas the polarization function with the
nucleonic part contains all the difficult physics related to nuclear
interactions in the dense and hot medium.
\subsection{Calculation of the polarization function}
\label{ss:polarization}
Within this section we will present the different approximations which
we have employed in order to calculate the polarization
function. First of all, we consider the nucleonic form factors being
constant neglecting any momentum dependence and corrections from
weak magnetism. The former is anyway very small since the energies in
our case are well below the relevant scale of
1~GeV~\cite{Horowitz:2001xf}. The latter
correction~\cite{Horowitz:2001xf} will be included in future
work. Second, since nuclear masses are much higher than typical
energies, in this work the non-relativistic approximation will be
employed. Let us mention that relativistic corrections might be
important, in particular if effective masses become of the same order
as other
energies~\cite{Leinson:2001ei,Leinson:2002bw,Roberts:2016mwj}, but a
closer inspection of this question will be kept for future
work. Applying these two assumptions, the polarization function can be
written as
\begin{equation}
  \Pi^R_{\ls} = g_{\lambda 0} g_{\sigma 0} g_V^2 \Pi_V + (g_{\lambda
    0} g_{\sigma 0} - g_{\ls}) g_A^2 \Pi_A~,
\end{equation}
with a vector contribution $\Pi_V$ and an axial one $\Pi_A$. $g_{V/A}$
are the nucleonic (axial) vector form factors. The metric $g^{\ls}$
denotes here the flat space Minkowski metric with signature
$(1,-1,-1,-1)$. From vector current conservation $g_V = 1$, and
$g_A/g_V = 1.2695$ from free neutron decay. A contraction with the
lepton tensor, see Eq.~(\ref{eq:ltensor}) then yields
\begin{align}
  L^{\ls} \Pi^R_{\ls} &= 8 \left(\Pi_V ( 2 E_e E_\nu - K_e \cdot K_\nu)\right. \nonumber \\ &\qquad{} \left. + \Pi_A ( 2 E_e E_\nu + K_e\cdot K_\nu)\right)~.
\end{align}

\subsubsection{Elastic approximation}
Many works in the literature consider the so-called elastic approximation, where the momentum transfer to the nucleons is neglected~\cite{Bruenn:1985en}. In that case we can write
\begin{align}
  \mathrm{Im}\, \Pi_V(q) & = \mathrm{Im}\,\Pi_A(q) \nonumber \\ &= - \pi \, (n_p - n_n) \, \delta(\tilde{q_0} + m_p - m_n)~,
\label{eq:lindhardelastic}
\end{align}
where $n_{n/p}$ are the neutron/proton number densities and $m_{n/p}$
the neutron/proton masses, respectively, and the energy argument of
the $\delta$-function, $\tilde{q_0}$ is shifted with respect to $q_0$
by the difference in proton and neutron chemical potentials,
$\tilde{q_0} = q_0 + \mu_n - \mu_p$. The remaining integration over
electron momenta in Eq.~(\ref{eq:dfnudt}) can then be carried out
analytically. We obtain
\begin{align}
\frac{\partial }{\partial t} f_\nu &=  \frac{G_F^2 V_{ud}^2}{\pi} \, (g_V^2 + 3 g_A^2) \sqrt{1 - \frac{m_e^2}{E_e^2}} E_e^2  (n_p - n_n)\nonumber \\ &\times \left((1-f_\nu) f_F(E_e-\mu_e) ( 1 + f_B(q_0))\right. \nonumber \\ & \qquad{} \left. -f_\nu(1- f_F(E_e-\mu_e)) f_B(q_0)\right)~,
\label{eq:dfnudtelastic}
\end{align}
with $E_e = E_\nu + m_n - m_p - (\mu_n - \mu_p - \mu_e + \mu_\nu) =
E_\nu + m_n - m_p$ and $q_0 = m_n - m_p + \mu_p - \mu_n$. This
expression is in agreement with the result in
\textcite{Bruenn:1985en}. The corresponding expressions for
anti-neutrino and positronic reactions can be obtained analogously,
explicit formulas are listed for completeness in
appendix~\ref{app:opacities}. Corrections to these expressions, taking
into account the phase space to first order in $E_\nu/m_i$, have been
derived in \cite{Horowitz:2001xf}.
\subsubsection{Mean field approximation}
It has been pointed
out~\cite{Reddy:1997yr,Roberts:2012um,MartinezPinedo:2012rb} that mean
field corrections to charged current processes can become important in
asymmetric matter such as in CCSN or neutron stars and BNS mergers
since the neutron and proton energy differences are enhanced by a
difference in mean field interaction potentials. In mean field, the
interaction can be recast into the definition of effective masses,
chemical potentials and/or single particle energies of the nucleons,
such that formally the system can be treated as a
free gas, with additional self-consistent equations determining the
effective quantities and potential terms for energy and pressure. In
particular the distribution function still has the form of a free
gas. For instance, in relativistic mean field models
\begin{equation}
  f_F(\sqrt{\vec{k}^2 + m_i^2} - \mu_i) \to f_F(\sqrt{\vec{k}^2 + (m_i^{*})^2} - \mu^*_i)
\end{equation}
with effective masses $m_{i}^*$ and chemical potentials $\mu_{i}^*$
and the index i stands for neutrons or protons,
respectively. Considering the momenta being much smaller than the
masses, we reach the non-relativistic limit considered here with
\begin{equation}
  \frac{\vec{k}^2}{2 \,m_i} + m_i - \mu_i \to \frac{\vec{k}^2}{2 \, m_i^*} + m_i^* - \mu_i^*~.
  \label{eq:effective}
\end{equation}
Eq.~(\ref{eq:effective}) can be rewritten in terms of non-relativistic mean field interaction potentials $U_i = \mu_i - \mu_i^*$ as~\cite{Hempel:2014ssa}
\begin{equation}
  \frac{\vec{k}^2}{2 \,m_i} + m_i - \mu_i \to \frac{\vec{k}^2}{2 \, m_i^*} + m_i^* + U_i - \mu_i~.
  \end{equation}
which resembles the standard definition of interaction potentials in non-relativistic Skyrme models, see e.g.~\cite{Ducoin:2005aa},
\begin{equation}
  \frac{\vec{k}^2}{2\,m_i} + m_i - \mu_i \to \frac{\vec{k}^2}{2\,m_i^{*}} + m_i +  U_i^{Sky} - \mu_i~.
\end{equation}
For later convenience we will introduce a common notation
$\epsilon_k^i = \frac{\vec{k}^2}{2 \, m_i^*} + m_i^*$, defining an
effective chemical potential for Skyrme models as
$\mu_i^*(\mathrm{Skyrme}) = \mu_i - U_i^{Sky} + m_i^* - m_i = \mu_i
- U_i$. Note the additional difference between the free and the
effective mass which arises from the different definitions of the
effective mass in relativistic and non-relativistic models.

The exact values of these effective quantities depend of course on the
equation of state, but it should be noted that the difference in
proton and neutron potentials can reach several tens of MeV in
asymmetric matter. Note in addition that the calculations of the
interaction potentials within the virial expansion in
\textcite{Horowitz:2012us} suggest that the potential difference
between protons and neutrons is underestimated by mean field
calculations.

These
corrections can be incorporated into the rates from the elastic
approximation: the nucleon masses in Eq.~(\ref{eq:lindhardelastic})
become effective masses $m_i \to m_i^*$ and $\tilde{q_0}$ becomes $\tilde{q_0} = q_0 + \mu_n^* - \mu_p^*$ meaning in particular that there is an additional shift by $U_p - U_n$. Eq.~(\ref{eq:dfnudtelastic}) then becomes
\begin{align}
\frac{\partial }{\partial t} f_\nu &=  \frac{G_F^2 V_{ud}^2}{ \pi} \, (g_V^2 + 3 g_A^2) \sqrt{1 - \frac{m_e^2}{E_e^2}} E_e^2 \nonumber \\ &  \left( (1-f_\nu) f_F(E_e -\mu_e)\eta_{pn}\right.\nonumber\\ & \qquad{} \left.-f_\nu(1- f_F(E_e-\mu_e)) \eta_{np}\right)~,
\label{eq:dfnudtelastic_mf}
\end{align}
where $E_e = E_\nu + m_n^* - m_p^* + U_n - U_p$ and
\begin{equation}
\eta_{ij} = \frac{n_i - n_j}{1 - \exp((-m_j^* + m_i^* - U_j + U_i - \mu_i + \mu_j)/T)}~.
\end{equation}
This result is in agreement with the expression in
\textcite{Bruenn:1985en}, modified due to mean field effects, see
e.g. \textcite{Fischer:2018kdt}. Again, explicit formulas for
anti-neutrino and positron reactions can be found in
appendix~\ref{app:opacities}.

The next step is to relax the elastic approximation, i.e. to include
the full nucleonic phase space. Staying on the mean field level, the polarization function becomes
\begin{equation}
  \mathrm{Im}\, \Pi_V(q)  = \mathrm{Im}\,\Pi_A(q) = 2 \mathrm{Im} \, L(q)~,
\label{eq:lindhardfull}
\end{equation}
with the well-known Lindhard function $L(q)$,
\begin{equation}
L(q) = \lim_{\eta \to 0} \int \frac{d^3 k}{( 2 \pi)^3}\frac{f_{F}(\ep - \mu_p^*) - f_F(\en-\mu_n^*)}{\tilde{q_0} + i \eta  + \ep - \en}~.
\label{eq:lindhard}
\end{equation}
An analytic expression can be derived for its imaginary
part~\cite{Reddy:1997yr}, see appendix~\ref{app:lindhard}. Please note
that the expression proposed in \textcite{Burrows:1998ek} neglects the
difference in effective masses between protons and neutrons --which
can be important in asymmetric matter --, and does not consistently
include the effect of nucleonic interaction potentials, see the
discussion of that point in \textcite{Roberts:2012um}, too.
The
integration over electron momenta in Eq.~(\ref{eq:dfnudt}) to obtain
the opacities can, in contrast to the elastic approximation, no longer
be performed analytically. In section~\ref{ss:results} we will discuss
numerically calculated opacities from this mean field approach with
full phase space.
\subsubsection{RPA polarization function}
\label{ss:rpa}
Technically, in order to obtain the RPA polarization function, in the
above expressions for the neutrino and anti-neutrino rates, the
Lindhard function has to be replaced by the RPA vector and axial
polarization function, respectively. Previous
works~\cite{Reddy:1998hb,Burrows:1998ek} have considered the so-called
Landau approximation, where the polarization function can be written
as~\cite{Hernandez:1999zz}
\begin{align}
\mathrm{Im} \,\Pi_V(q) &=\mathrm{Im}\, \frac{L(q)}{1 - 2 f_{cc} L(q)}~,\label{eq:landauv} \\
\mathrm{Im} \,\Pi_A(q) &=\mathrm{Im}\, \frac{L(q)}{1 - 2 g_{cc} L(q)}~,
\label{eq:landaua}
\end{align}
where $f_{cc}$ and $g_{cc}$ represent the residual interaction and the
factor two arises from spin degeneracy. Explicit expressions for the
parameters $f_{cc}$ and $g_{cc}$ in terms of the usual Skyrme
parameters can be found in \textcite{Hernandez:1999zz}, extended to
asymmetric matter~\cite{Margueronphd}. Note that these Landau
parameters depend on the Fermi momenta of protons and neutrons
separately and are thus density and charge fraction dependent in
contrast to the constant symmetric matter values assumed in
\textcite{Burrows:1998ek}, see the criticism in
\textcite{Horowitz:2006pj}. A relativistic version can be found in
\textcite{Reddy:1998hb}. Please note that in Eq.~(\ref{eq:landaua})
the real part of the Lindhard function enters, too. For equal masses,
and assuming additionally the classical (Boltzmann) limit for the
distribution functions, an analytic expression for this real part
can be obtained~\cite{Burrows:1998cg}.  This is no longer possible
including the full mean field effects with effective masses and
interaction potentials and applying Fermi-Dirac statistics for the
nucleons. Therefore, we compute the real part via a dispersion
relation from Eq.~(\ref{eq:imlim}).

Unfortunately, most of the Skyrme forces show an instability in the
spin-isospin (axial) channel at high density~\cite{Pastore:2014aia},
leading to a diverging (anti-) neutrino opacity. This channel is
anyway very badly constrained by the usual fitting procedure of Skyrme
forces. Two possible remedies have been proposed in the literature:
\begin{enumerate}
\item Employ a microscopically motivated residual interaction in this channel instead of Eq.~(\ref{eq:landaua}), see e.g.~\citet{Reddy:1998hb}, Eqs.~(53)-(57), or \cite{Rapp:1997ei}. The axial response then becomes
  \begin{equation}
    \mathrm{Im} \,\Pi_A(q) =\mathrm{Im}\,L(q) \left( \frac{1}{3 D_L(q)} + \frac{2}{3 D_T(q)}\right)~,
  \end{equation}
  with a transverse $D_T$ and a longitudinal $D_L$ part,
  \begin{align}
    D_I(q) &= \{1 - 2 V_I(q) \mathrm{Re}\, L(q)\}^2\nonumber\\ & \qquad{} + \{ 2 V_I(q)\mathrm{Im}\, L(q)\}^2~.
    \end{align}
  The residual interaction is given by
  \begin{align}
    V_L(q) &= \frac{f_{\pi NN}^2}{m_\pi^2} \left(\frac{\vec{q}^2}{q^2 - m_\pi^2} F_\pi^2(q) + g'\right) \\
    V_T(q) &= \frac{f_{\pi NN}^2}{m_\pi^2} \left(\frac{2 \,\vec{q}^2}{q^2 - m_\rho^2} F_\rho^2(q) + g'\right) ~.
    \label{eq:vlt}
    \end{align}
  The $\pi NN$ and $\rho NN$ form factors are taken as $F_\pi = (\Lambda_\pi^2 - m_\pi^2)/(\Lambda_\pi^2 - q^2)$ and $F_\rho = (\Lambda_\rho^2 - m_\rho^2)/(\Lambda_\rho^2 - q^2)$. For numerical applications, we will take for the parameters~\cite{Rapp:1997ei}: $f_{\pi NN} = 1.01, g'= 0.6, m_\pi = 140 $ MeV, $m_\rho = 770$ MeV, $\Lambda_\pi = 550$ MeV, $\Lambda_\rho = 1$ GeV.

  As can be seen from Eq.~(\ref{eq:vlt}), in this case the residual
  interaction becomes momentum dependent and is suppressed for high
  momenta $|\vec{q}|$, as it is expected from microscopic
  calculations.
  \item Add an additional repulsive term in this particular channel,
    without changing the remaining properties of the model and in
    particular the equation of state~\cite{Margueron:2009rn}. In this
    case, $g_{cc} \to g_{cc} + t_3'\, n_B^2/4$ in
    Eq.~(\ref{eq:landaua}) with the additional parameter $t_3'$ which
    will be taken as $t_3'= 1\times 10^4$ MeV$\cdot$ fm$^9$\footnote{This value is half the value of
      \textcite{Margueron:2009rn}, but it still garantuees stability
      at high densities and does only marginally change the results at
      low density since the correction term is $\propto n_B^2$.} in
    numerical applications~\cite{Margueron:2009rn} in numerical
    applications~\cite{Margueron:2009rn}. This approach has the
    advantage of remaining coherent with the underlying EoS.
\end{enumerate}

The formalism for the full RPA with contact Skyrme type interactions
in the charge exchange channel has been presented in
\textcite{Hernandez:1999zz} and extended to Skyrme forces with tensor
terms~\cite{Davesne:2019ytl}. Some first results for charged-current
neutrino opacities have recently been
discussed~\cite{Dzhioev:2018ovi}. However, we expect the essential
effect of RPA correlations on the neutrino opacities to be already
comprised in the Landau approximation, which is numerically much
faster to evaluate, at least as long as the momentum transfer does not
become too large --recall that it results from a low-momentum
expansion. In addition, the instability in the spin-isospin channel
mentioned above, although less pronounced in full RPA due to the
momentum dependence of the residual interaction, persists with a
diverging opacity already for densities and temperatures relevant for
(proto-)neutron star and post-merger matter for many standard Skyrme
forces. We have therefore decided to produce the complete opacity data
within Landau approximation, adding the repulsive term from
\textcite{Margueron:2009rn} in the spin-isospin channel to keep
consistence with the EoS.

\subsection{Resulting neutrino opacities}
\label{ss:results}
\subsubsection{Equation of state}
\label{ss:eos}
During the different stages of the core collapse evolution or during a
BNS merger wide domains of density ($10^{-12} \lesssim n_B \lesssim 1$
fm$^{-3}$), temperature ($ 0.1 \lesssim T \lesssim 50$ MeV) and charge
fraction ($ 0.01 \lesssim Y_e \lesssim 0.6$) are explored. Matter
composition changes throughout with nuclear clusters present at low
densities and temperatures and homogeneous matter elsewhere. Matter
consists of baryons --in the simplest case just nucleons and nuclear
clusters--, leptons and photons. Charged leptons and photons are
usually treated as ideal Fermi and, respectively, Bose gases, whereas
neutrinos, being in general not in equilibrium, are not included in the
EoS. The detailed composition and thermodynamics of baryonic matter is
still under debate, due to the uncertainties in effective interactions
and the difficulties in the modeling a strongly interacting many-body
system.

\begin{table*}[t]
  \squeezetable
  \begin{tabular}{c|c|c||c|c|c|c|c|c|c|c}
    \hline
    $T$& $n_B$ & $Y_e$ &EoS & $x_n$ & $x_p$ & $U_n$ & $U_p$ & $\Delta U$& $m_n^*$ & $m_p^*$  \\
    (MeV) & (fm$^{-3}$)& & & & & (MeV) & (MeV) &(MeV) & (MeV)& (MeV) \\ \hline
    \hline
5.25 & $10^{-4}$ & 0.1 & HS(DD2) & 0.89& 0.090 &0.372 &0.293 & 7.89$\times 10^{-2}$ & 939.1& 937.8 \\
5.11 & 1.15$\times 10^{-4}$ & 0.1 & RG(SLy4) & 0.89& 0.089 &0.312 &-0.113 &0.2202 & 939.2& 938.1 \\
19.1 & 5.25$\times 10^{-3}$ & 0.1 & HS(DD2) & 0.86& 0.069 &18.46 &14.48 & 3.979 &918.7& 917.4 \\
19.5 & 5.29$\times 10^{-3}$ & 0.1 & RG(SLy4) & 0.87& 0.072 &15.21 &0.794 &-6.842 & 922.3& 930.2 \\
5.25 & 1.20$\times 10^{-2}$  & 0.15 & HS(DD2) & 0.54& 0.008 & 25.93&19.39 &6.531 &910.0& 908.7 \\
5.11 & 1.19$\times 10^{-2}$ & 0.15 & RG(SLy4) & 0.35& 0.004 &13.50 &-2.032 & 6.642 & 924.5& 932.1 \\
12.0 & 1.20$\times 10^{-2}$ & 0.1 & HS(DD2) & 0.82& 0.043 &38.67 &30.05 & 8.619 &895.6& 894.3 \\
12.1 & 1.19$\times 10^{-2}$ & 0.1 & RG(SLy4) & 0.75& 0.033 &28.66 &-1.308 &11.96 &907.3& 924.0 \\
8.32 & 1.10$\times 10^{-1}$ & 0.05 & HS(DD2) & 0.95 & 0.05 & 284.5 & 239.4 & 45.10&611.9 & 610.6 \\
8.19 & 1.10$\times 10^{-1}$ & 0.05 & RG(SLy4) & 0.95& 0.05 &233.0 &58.10 & 43.77 &663.0& 792.8 \\
\hline
  \end{tabular}
  \caption{Effective masses, interaction potentials and fractions of
    protons $x_p$ and neutrons $x_n$ under the thermodynamic
    conditions for which the different opacities are shown in
    Figs.~\ref{fig:T8n1} - \ref{fig:ccsn}.
    $\Delta U = m^*_n - m^* _p
    + U_n - U_p - (m_n - m_p)$ is the shift in reaction threshold due
    to mean field effects, see Eq.~(\ref{eq:dfnudtelastic_mf}).
  }\label{tab:thermo}
\end{table*}
For this work, we will use two different EoS models. As a fiducial
case for which we will calculate neutrino opacities in RPA, we will
consider the NSE model of Raduta and
Gulminelli~\cite{Gulminelli_PRC_2015,Raduta_2019} named
``RG(SLy4)''. It employs the non-relativistic SLy4 \cite{SLy4} Skyrme
interaction for nucleons. Mean field and elastic approximation results
will be compared with the NSE model of~\cite{Hempel_NPA_2010} with the
DD2 \cite{DD2} relativistic mean field interaction for the nucleons
(``HS(DD2)''), see \textcite{Hempel_ApJ_2012}. Both EoS models fulfill
constraints from nuclear experiments and the neutron star maximum
mass~\cite{Demorest_10,Antoniadis_13,Arzoumanian:2017puf} and are in
reasonable agreement with theoretical ab initio determinations of the
low density neutron matter EoS, see e.g.~\cite{Oertel:2016bki} for a
discussion. In addition, the tidal deformability calculated from
RG(SLy4) falls within the 90\% confidence interval for
GW170817~\cite{abbott_17,abbott_18} and HS(DD2) is marginally
compatible.  The contribution from electrons, positrons and photons is
included.

\begin{figure*}
\includegraphics[width=.9\textwidth]{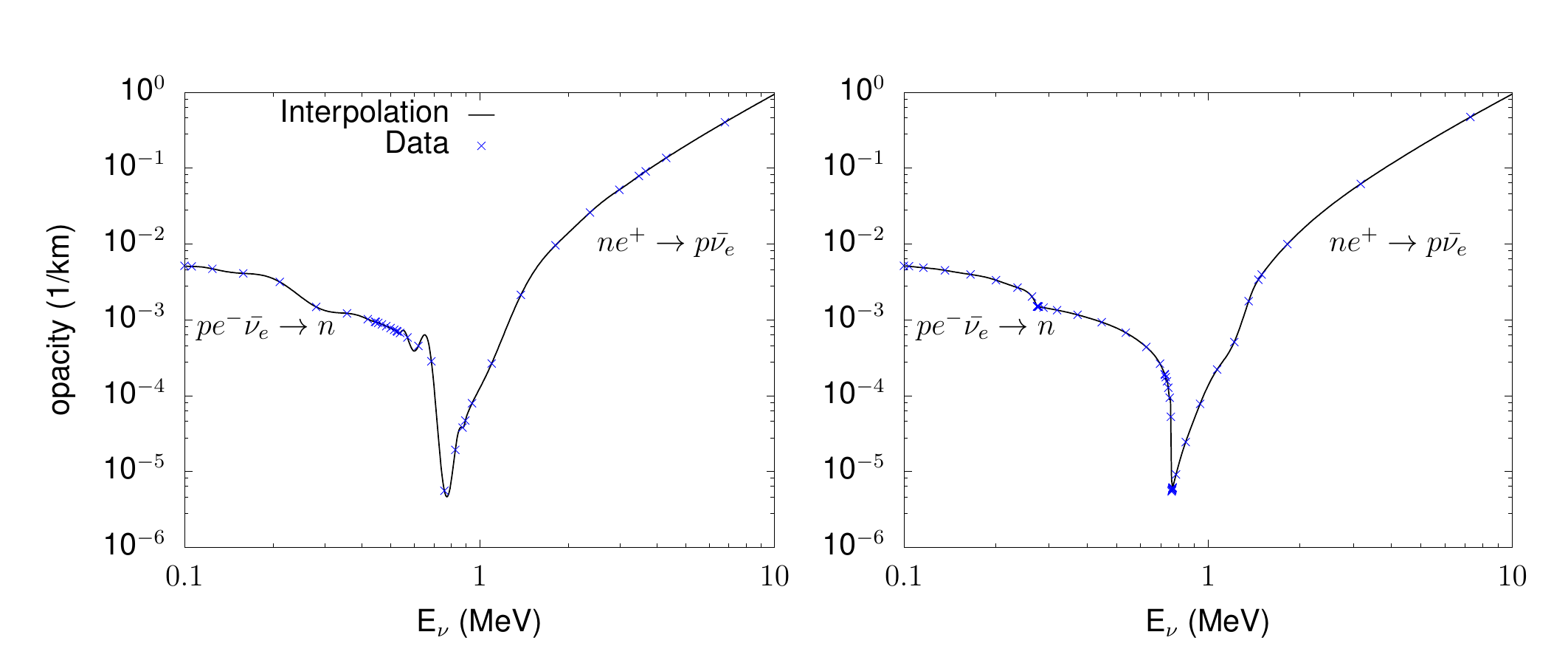}
\caption{(color online) Example of the interpolation procedure and
  distribution of the energy domains in the case of a pronounced
  reaction threshold for $T = 6$ MeV, $n_B = 0.025$ fm$^{-3}$, $Y_e$ =
  0.53 with RG(SLy4) EoS and RPA in Landau approximation. The crosses
  indicate the calculated points using five domains in energy, and the
  solid line corresponds to the interpolation. On the left hand side
  oscillations due to the Gibbs phenomenon in the interpolation
  function are clearly visible, whereas a better choice of the energy
  domains and thus a better distribution of the interpolated data
  points on the right hand side avoids them.
  \label{fig:interpolation}}
\end{figure*}

\subsubsection{Interpolation procedure}
\label{ss:interpolation}
For easy and fast evaluation of the neutrino opacities on the fly
during simulations, we interpolate the opacity data as functions of
neutrino energy and provide the coefficients for each grid point in
$(T, n_B, Y_e)$ of the corresponding EoS data. More precisely we
employ an eighth order polynomial for $\log(\kappa_a^*)$ and
$\log(\bar{\kappa_a^*})$ as functions of $\log(E_{\nu})$ to
interpolate the opacities for (anti-)neutrino energies $E_\nu$ between
0.1 and 250 MeV. An additional difficulty arises if the opacities show
rapid variation in a small energy interval, which can happen for
instance at the different reaction thresholds which become very
pronounced at low temperature. In order to avoid oscillations in the
interpolation due to the Gibbs phenomenon in this case, the entire
energy range $0.1 \le E_\nu \le 250$ MeV is divided into several
domains, where the interpolation prescription, see
Eq.~(\ref{eq:interpolation}) below, is applied in each domain. The
number of domains ($n_d$) and the domain borders $E_\nu^{\mathit{min/max}}$
are determined from the position of the thresholds and the variation
of the opacities in the vicinity of the respective thresholds.

The energy interval
$E_\nu \in [E_\nu^{\mathit{min}}, E_\nu^{\mathit{max}}]$ has been
mapped to the interval $\xi \in [-1,1]$ via an affine mapping
$\log(E_\nu) = \alpha \xi + \beta$. The opacities are then computed
via
\begin{equation}
  \log(\kappa) = \sum_{n=0}^N c_n(T,n_B,Y_e) \xi^n~
  \label{eq:interpolation}
\end{equation}
with the coefficients $c_n$ depending on the thermodynamic conditions and $N = 8$.

\begin{figure*}
\includegraphics[width=.9\textwidth]{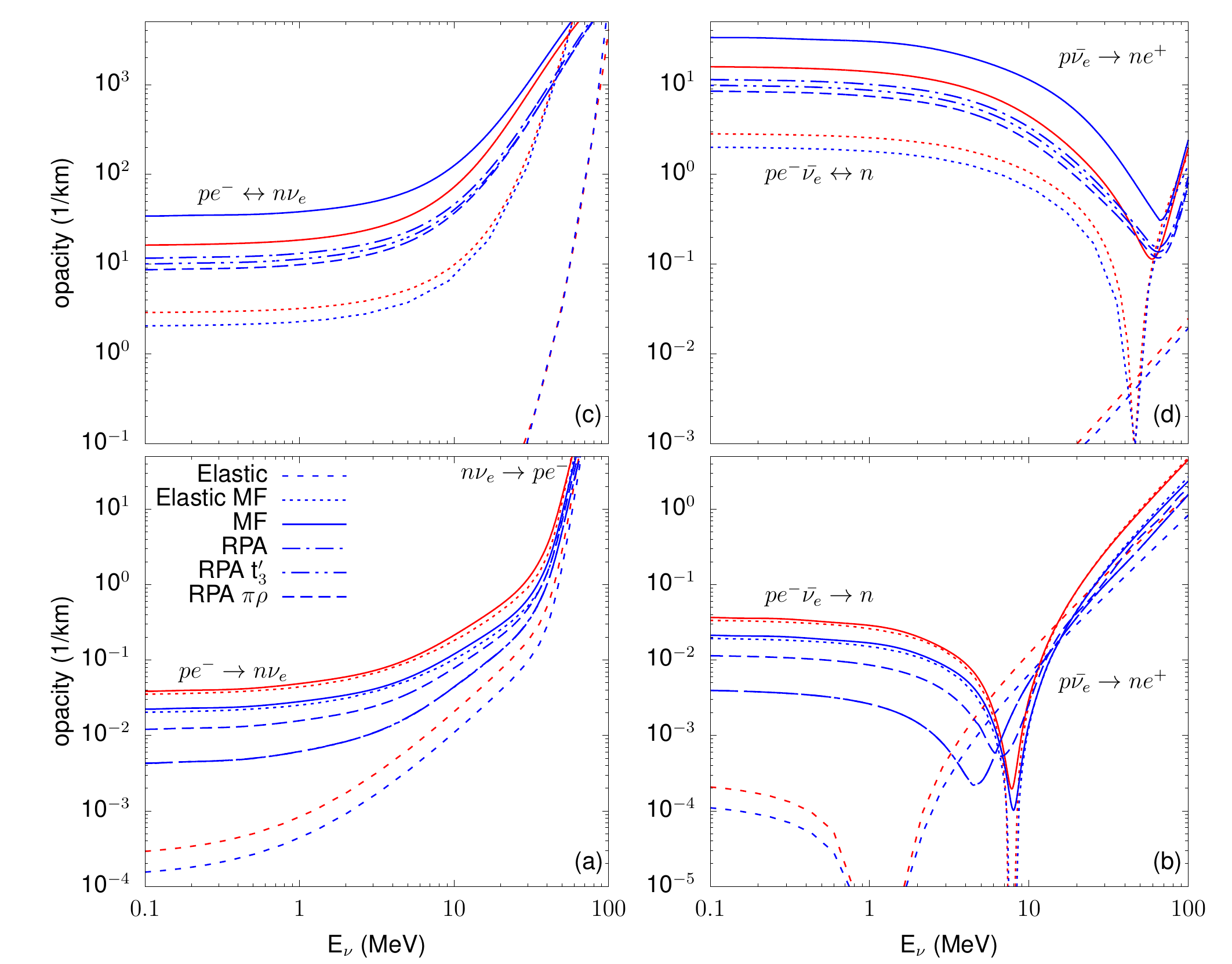}
\caption{(color online) Neutrino (left) and anti-neutrino (right)
  opacities for $T = 8$ MeV, $n_B = 0.11$ fm$^{-3}$ and $Y_e = 0.05$
  (upper panels), and $T = 5$ MeV, $n_B = 0.01$ fm$^{-3}$ and $Y_e =
  0.15$ (lower panels). These correspond to typical conditions for
  the decoupling of neutrinos in the merger
  remnant~\cite{Endrizzi:2019trv}, the former are more relevant for
  lower neutrino energies than the latter. The different line types
  distinguish the different approximations and results with HS(DD2)
  are indicated in red, whereas those with RG(SLy4) are in blue, see
  section~\ref{ss:polarization} for details. The dominant processes
  contributing to the opacities in a certain energy domain are
  mentioned in the figure, too.
    \label{fig:T8n1}}
\end{figure*}

Fig.~\ref{fig:interpolation} shows an example of the choice of domains
to avoid an oscillating interpolation function in the case of a
threshold. It corresponds to $T = 6$ MeV, $n_B = 0.025$ fm$^{-3}$,
$Y_e$ = 0.53 and the opacities have been calculated with the RG(SLy4)
EoS employing RPA in Landau approximation with a nonzero $t_3'$. At
low $E_\nu$ the reaction $p + e^- + \bar{\nu} \to n$ is dominant,
whereas at higher energies $n + e^+ \to p + \bar{\nu}$ overtakes. The
threshold slightly below $E_\nu = 1$ MeV is clearly visible and the
opacity varies by orders of magnitude in a narrow energy interval
close to this threshold. The oscillations in the interpolation on the
left panel are clearly visible, whereas a better choice of domain
border reduces them considerably, see the right panel.
\begin{figure*}
\includegraphics[width=.9\textwidth]{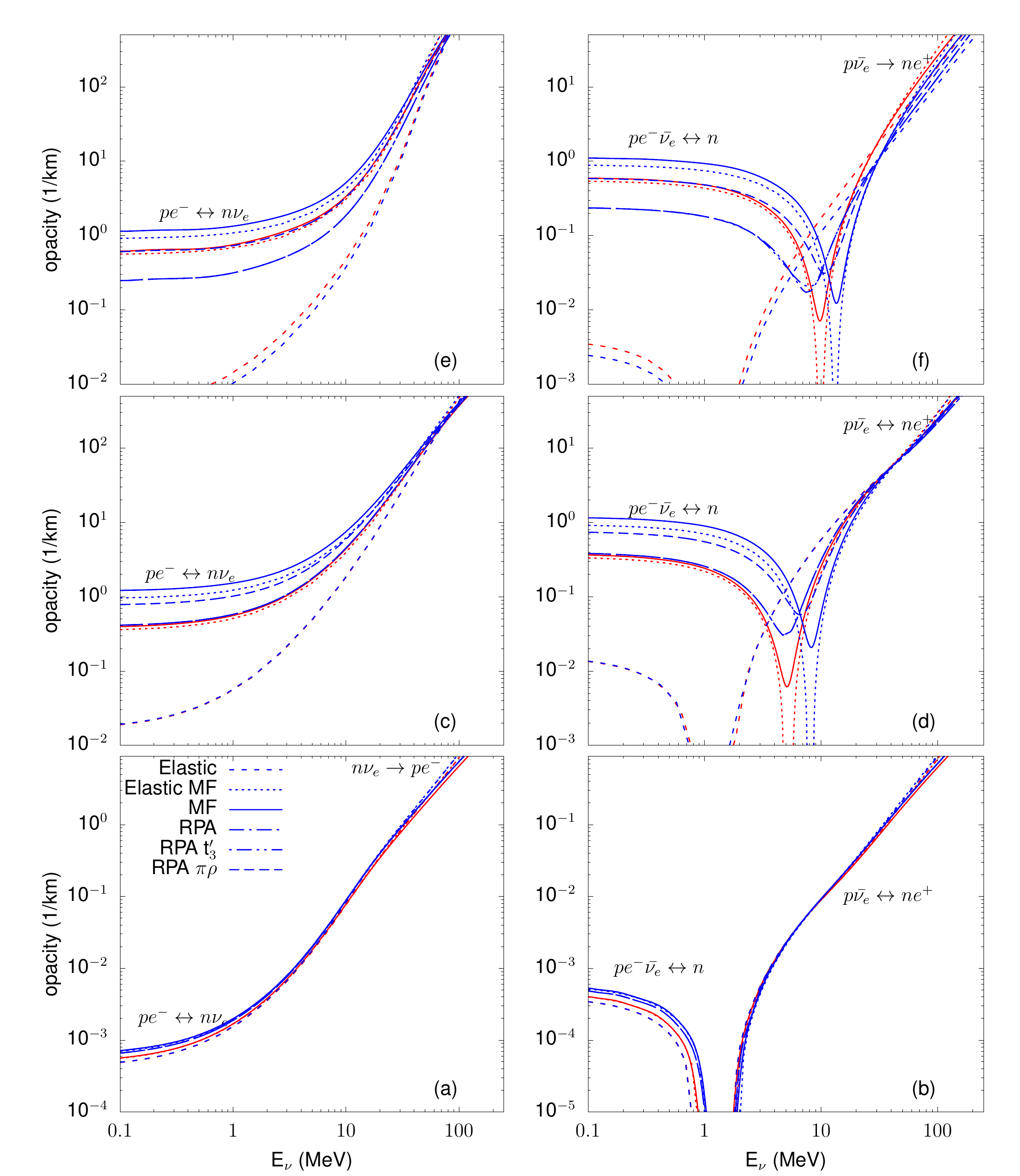}
\caption{(color online) Neutrino (left) and anti-neutrino (right)
  opacities for $T = 12$ MeV, $n_B = 0.01$ fm$^{-3}$ and $Y_e = 0.1$
  (upper panels), $T = 19$ MeV, $n_B = 5\times 10^{-3}$ fm$^{-3}$ and
  $Y_e = 0.1$ (middle panels) and $T = 5$ MeV, $n_B = 10^{-4}$
  fm$^{-3}$ and $Y_e =0.1$ (lower panels). These correspond to
  typical conditions close to the neutrinosphere in a CCSN from our
  fiducial simulation, see section~\ref{s:ccsn}. The different
  line types distinguish the different approximations and results with
  HS(DD2) are indicated in red, whereas those with RG(SLy4) are in
  blue. The dominant processes contributing to the opacities in a
  certain energy domain are mentioned in the figure, too.
    \label{fig:ccsn}}
\end{figure*}
\subsubsection{Opacities}
\label{ss:opacities}
Let us now discuss the opacities resulting from the different
approximations for various thermodynamic conditions. The most
interesting regions are probably located close to the respective
neutrinospheres. Since the
opacities depend on neutrino energy and flavor, it is obvious that the
location where neutrinos decouple from matter varies as function of
these quantities. It has been observed that low energy neutrinos have
in general lower opacities, i.e. longer mean free paths and decouple
further inside and thus at higher densities and temperatures than high
energy neutrinos.

A comprehensive analysis for binary merger simulations indicates
densities $10^{-4} \lesssim n_B \lesssim 10^{-1}$ fm$^{-3}$ ,
temperatures $T \sim$ 1-10 MeV and electron fractions $Y_e\sim$ 0.05 -
0.3 at the corresponding neutrinospheres for neutrino energies $E_\nu
\sim$ 3-100 MeV~\cite{Endrizzi:2019trv}. The highest densities and
temperatures as well as the lowest electron fractions are thereby
associated with the lowest neutrino energies. Although evolving with
time and being sensitive to the simulation setup, in particular the
neutrino treatment and the EoS, these values can be considered as
typical ones for a binary merger remnant. Fig.~\ref{fig:T8n1} shows
the opacities employing the different approximation schemes for two
different thermodynamic conditions, chosen within the above
ranges. The upper panels, with $T = 8$ MeV, $n_B = 0.11$ fm$^{-3}$ and
$Y_e = 0.05$ are more relevant for neutrinos with low energies,
whereas the lower panels show $T = 5$ MeV, $n_B = 0.01$ fm$^{-3}$ and
$Y_e = 0.15$, conditions close to decoupling for neutrinos with
slightly higher energies.

The passage of the shock heats up matter between the proto-neutron
star surface and the neutrinosphere in a CCSN such that compared with
the conditions of the binary merger remnant, for CCSN we have to
consider slightly higher temperatures with very similar densities and
electron fractions. From our fiducial simulations, see
section~\ref{s:ccsn}, we have chosen three different thermodynamic
conditions for which opacities are displayed in Fig.~\ref{fig:ccsn}:
$T = 12$ MeV, $n_B = 0.01$ fm$^{-3}$ and $Y_e = 0.1$ (upper panels),
$T = 19$ MeV, $n_B = 5\times 10^{-3}$ fm$^{-3}$ and $Y_e = 0.1$
(middle panels) and $T = 5$ MeV, $n_B = 10^{-4}$ fm$^{-3}$ and $Y_e
=0.1$ (lower panels). The neutrinosphere for $\bar{\nu}_e$ is thereby
located slightly closer to the center, i.e. at slightly higher
densities and temperatures. The first example (upper panels) thereby
correspond roughly to conditions at the neutrinosphere for low energy
$\bar{\nu}_e$ at early post-bounce and for $\nu_e$ at later times, the
second is relevant for low energy anti-neutrinos and the third for
both $\nu_e$ and $\bar{\nu}_e$, but with higher energies of the order
ten MeV. The neutrinospheres of (anti-)neutrinos with still higher energies
are located at lower densities and temperatures.

As can be seen from Fig.~\ref{fig:ccsn}, lower panels, at $n_B =
10^{-4}$ fm$^{-3}$, the difference between the approximation schemes
is very small, the largest difference is reached for $E_\nu \lesssim
1$ MeV and does not exceed a factor 1.5. There are two reasons for
that: first, only small momentum transfers are involved, such that the
elastic approximation \footnote{Note that the difference in neutron
  and proton number densities entering Eq.~(\ref{eq:dfnudtelastic})
  should in principle be calculated with free masses and chemical
  potentials, differing thus from the values given by the EoS. For the
  curves labeled ``elastic'', we have employed the densities from the
  EoS, thus including already some mean field effects. } is well
justified for the present conditions; second, as well mean field as
RPA effects arise due to interactions in the dense nuclear medium and
the corresponding corrections are thus small at low densities. In
Table~\ref{tab:thermo} we list effective masses and interaction
potentials for the thermodynamic conditions of Figs.~\ref{fig:T8n1}
and \ref{fig:ccsn} and the two employed EoS and it can easily be
checked that the interaction potentials are indeed small and effective
masses are close to the free masses in the present case.

The situation becomes different at higher densities. At
$n_B = 5\times 10^{-3}$ fm$^{-3}$, see panel (c) and (d) in
Fig.~\ref{fig:ccsn}, momentum transfer is still small for the
considered neutrino energies, such that the mean field results agree
well between the elastic approximation (dotted lines) and the full
phase space integration (solid lines) for both EoS. The only
noticeable difference is that the opacities do not vanish any more
close to the reaction thresholds upon full phase space integration,
see panel (d). The mean field corrections, however, are large. The
most prominent effect is the shift of the reaction threshold by
$\pm \Delta U = -( m^*_n - m^* _p + U_n - U_p - (m_n - m_p))$ where
the lower sign corresponds to neutrinos and the upper one to
anti-neutrinos, respectively. This shift is clearly visible for the
anti-neutrino opacities and it is more pronounced for RG(SLy4) since
$|\Delta U|$ is larger, see Table~\ref{tab:thermo}. For neutrinos, no
reaction threshold lies within the shown energy range. RPA
correlations tend to decrease the shift in reaction threshold and push
it to lower anti-neutrino energies. Both, neutrino and anti-neutrino
opacities are strongly suppressed for low $E_\nu \lesssim 1$ MeV and
approach the mean field results at higher energies. In particular, for
$E_\nu \gtrsim 30$ MeV almost no difference is observed any more. This
clearly shows, together with the shift in reaction thresholds, that
RPA correlations cannot be cast into a grey correction factor,
i.e. multiplying the mean field rates by a common factor for all
(anti-)neutrino energies.

The prevailing role of the
axial (spin-isospin) channel in the RPA results, see
e.g. \textcite{Reddy:1998hb}, is confirmed by the shown opacities. The
vector channel is treated in the same way for all RPA models, The
observed non negligible difference in the opacities is thus entirely
due to the different prescriptions chosen for the axial channel, see
section~\ref{ss:rpa}. In all figures ``RPA'' denotes the results
obtained by employing the $g_{cc}$ parameter from the standard Skyrme
interaction~\cite{Hernandez:1999zz}, ``RPA $t_3^\prime$'' those with
an additional repulsive term~\cite{Margueron:2009rn} and ``RPA
$\pi\rho$'' employs the microscopically motivated
$\pi\rho$-model~\cite{Reddy:1998hb}. For the present case at $T = 19$
MeV, $n_B = 5\times 10^{-3}$ fm$^{-3}$ and $Y_e = 0.1$, the $\pi\rho$
model leads to a suppression by roughly a factor 1.5 with respect to
mean field results at low $E_\nu$, whereas for the two other RPA
models the suppression is about twice as strong and reaches roughly a factor
three. Similarly large differences following the prescription for the
residual interaction in the axial channel are seen for other
thermodynamic conditions, see Figs.~\ref{fig:T8n1} and
\ref{fig:ccsn}. The uncertainties in the opacities due to this badly
constrained interaction probably predominate over the differences
between Landau approximation and full RPA, although the latter might
become important at high densities with higher momentum transfers.

At these conditions the difference in opacities between both EoSs
using the mean field approximation is almost as important as the
differences between RPA and mean field results. This is no longer the
case at $n_B = 10^{-2}$ fm$^{-3}$, see Fig.~\ref{fig:ccsn}, panels (e)
and (f) as well as Fig.~\ref{fig:T8n1}, panels (a) and (b). At low
$E_\nu$ opacities in RPA are suppressed by about a factor five at
$T = 5$ MeV, $Y_e = 0.15$ and up to a factor ten at $T = 12$ MeV and
$Y_e = 0.1$, whereas the mean field results of RG(SLy4) and HS(DD2)
differ only by about a factor two. The shift in the anti-neutrino
reaction threshold is more pronounced for RPA, too. The opacities at
higher $E_\nu$ again become very similar within all the different
approximations employed. Please note that at neutrino energies above
those shown here, momentum transfer becomes high enough to induce
again noticeable differences between the results with full phase space
integration and the elastic ones. At still higher densities, see
Fig.~\ref{fig:T8n1} (upper panels), where opacities are displayed for
$n_B = 0.11$ fm$^{-3}$, $T=8$ MeV and $Y_e = 0.05$, qualitatively the
behavior is rather similar to the previously discussed
cases. Quantitatively, the effect of mean field corrections increases
as expected and the momentum transfer becomes higher such that the
elastic mean field results no longer reproduce well the full phase
space integration. The mean field calculations of
\textcite{Roberts:2016mwj} using relativistic kinematics show a
similar trend: the full space integration becomes more important at
higher densities when the momentum transfer becomes larger. For the
particular case considered here, the three prescriptions for the
residual interaction in RPA Landau approximation lead to very similar
results. This should, however, not be seen as a general trend, but as
a result only valid for some particular thermodynamic conditions.

The dominant processes contributing to the charged-current opacities
are indicated in Figs.~\ref{fig:T8n1} and \ref{fig:ccsn},
too. Generally simulations only consider the electron and positron
capture reactions as well as their inverse to compute opacities. As
already noticed in \textcite{Fischer:2018kdt}, for low energy
anti-neutrinos, (inverse) neutron decay becomes, however, the dominant
charged current process under these typical thermodynamic
conditions. These can even dominate over other opacity sources for low
energy anti-neutrinos such as for example $NN$-Bremsstrahlung,
customarily included in simulations~\cite{Fischer:2018kdt}. Let us
emphasize that the opacities computed here and the data provided
contain all different types of charged current reactions for electron
(anti-)neutrinos.

From the results for the opacities discussed here, we expect for CCSN
and BNS mergers that the properties of (anti-)neutrinos with an energy
of tens to several tens of MeV are only slightly modified, whereas low
energy (anti-)neutrinos experience more pronounced modifications. This
means in particular that the resulting spectra and luminosities should
be modified, too. As mentioned above, for (anti-)neutrinos with
energies above those shown and discussed here, differences in
opacities are expected to be due to the full phase space
integration. On the one hand, within a CCSN those are not very
numerous, and on the other hand their mean free path is extremely
small due to the very high Fermi energy of the electrons involved in
the different processes. Therefore, we will not discuss the detailed
effect on their spectra here.

\section{Core-collapse and early proto-neutron star evolution}
\label{s:ccsn}

In this section we discuss some first results implementing the newly
calculated opacities into a simulation for the early post-bounce
evolution in a CCSN. These results have been obtained using a
spherically-symmetric version of the \textsc{CoCoNuT} code
\cite{coconut}. This code solves the general-relativistic
hydrodynamics with a 3+1 decomposition of spacetime. High-resolution
shock-capturing schemes are used for hydrodynamic equations, whereas
Einstein equations for the gravitational field are solved with spectral
methods~\cite{dimmelmeier-05}. Energy losses and deleptonization via
neutrino interactions are computed using the "Fast Multi-group
Transport" (FMT) scheme~\cite{mueller-15}, which solves the stationary
neutrino transport problem using estimates of the flux factor obtained
by a two stream approximation in the optically dense region and an
Eddington factor closure in the optically thin region.

Both neutrino neutral and charged currents on nuclei, along with
neutral current scattering on nucleons, are considered with standard
opacities in the elastic approximation \cite{Bruenn:1985en} including the ion
screening effect \cite{Horowitz:1996ci}, whereas charged current
neutrino nucleon opacities are the subject of this work and vary throughout
the different simulations. Neither pair production reactions nor
inelastic scattering have been included for electron-flavor
(anti-)neutrinos. Anyway, for electron-flavor neutrinos, in the denser
area of the star the medium opacity is largely dominated by charged
current processes on nucleons/nuclei and the omitted reactions only
play a role in equilibration of the spectrum.
Heavy flavor neutrinos are treated as in \cite{mueller-15}.

The simulations start from an unstable stellar model taken among the
publicly available data published by \citet{woosley-02}. All results
presented in this section have been obtained using the \texttt{s15}
($15 M_{\odot}$ with solar metallicity) initial model, but we obtained
similar conclusions by testing other progenitors, in particular with
\texttt{u18} ($18 M_\odot$, $10^{-4}\times$ solar metallicity) and
\texttt{u40} ($40 M_\odot$, $10^{-4}\times$ solar metallicity)
progenitor models.

\subsection{Pre-bounce deleptonization}

Except during the last few milliseconds before trapping, pre-bounce
deleptonization is dominated by electron captures on neutron-rich
nuclei and the bounce properties are affected by the uncertainties on
the associated rates \cite{Hix2003,sullivan-16,pascal-20}. On the
contrary, only small differences for the electron fraction at bounce
are expected between different prescriptions for charged-current
reactions on free nucleons.

\begin{figure}[h]
\includegraphics[width=\columnwidth]{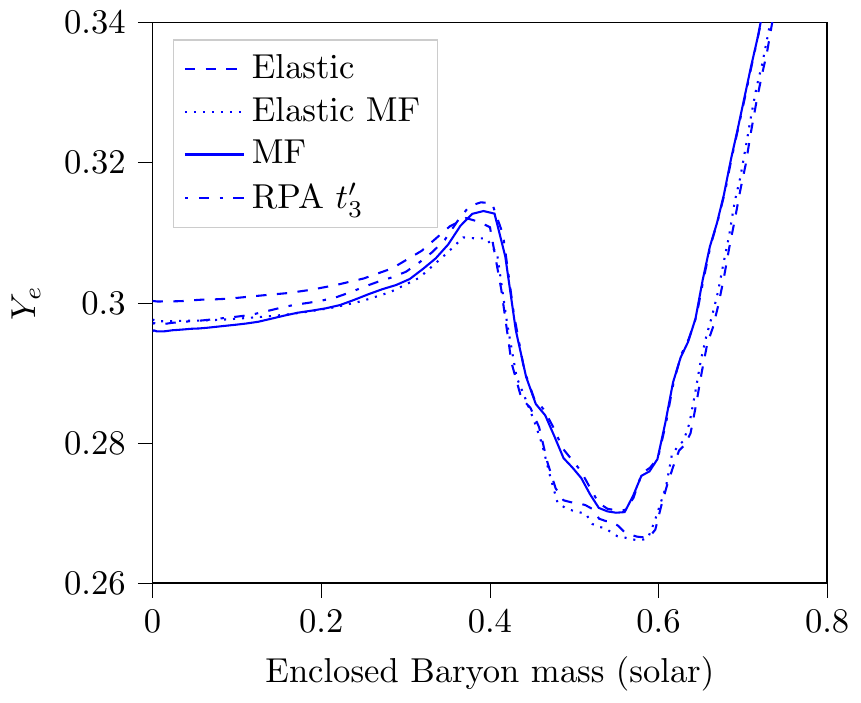}
\caption{(color online) Profiles of the electron fraction at bounce as
  a function of the enclosed baryon mass,
  employing different approximation schemes to compute charged-current
  neutrino-nucleon interaction rates.
  \label{fig:ye_bounce}}
\end{figure}

Indeed, the electron fraction profiles at bounce time are plotted in
Fig.~\ref{fig:ye_bounce}, for the four different models of
charged-current neutrino-nucleon interaction rates. For better
readability, these profiles are plotted as functions of the enclosed
baryon mass (i.e. for a given radius, we consider the baryon mass
contained inside this radius), which allows to take into account
possible time shifts at bounce. Differences between these models
remain small, showing that these reaction rates have, indeed, a small
influence during the pre-bounce phase.

\subsection{Post-bounce evolution}

It should be stressed that our simulations have been performed in
spherical symmetry, therefore the post-bounce evolution does not
reflect the strong asymmetries observed in 3D simulations (see, e.g.,
\cite{hanke-13}) and we cannot realistically investigate the effect of
the improved rates on shock revival and post-bounce dynamics. Let us
concentrate, therefore, on illustrating qualitatively the expected
effects. As discussed in Section~\ref{ss:opacities}, the main
modifications should concern neutrino luminosities and spectra due to
the changes in opacities for low energy (anti-) neutrinos
($E_\nu \lesssim 10$ MeV).

\begin{figure}[h]
\includegraphics[width=\columnwidth]{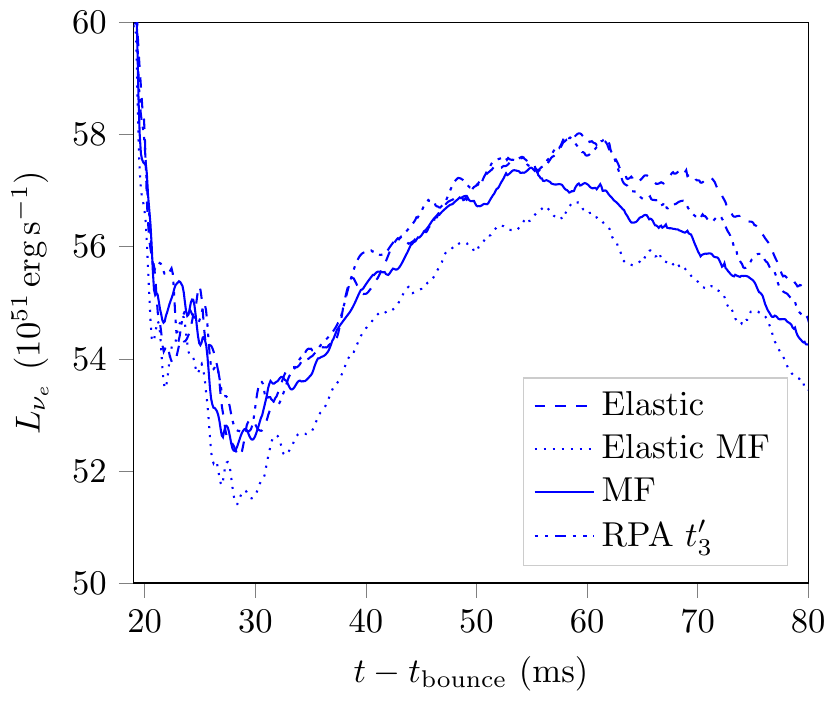}
\caption{(color online) Total neutrino luminosities as a function of
  time after bounce.
  \label{fig:nu_lum_pb}}
\end{figure}

The total early post-bounce luminosities for $\nu_e$ and $\bar{\nu}_e$,
with different prescriptions, as presented in Fig.~\ref{fig:nu_lum_pb}
(for $\nu_e$) are very similar. The reason is that at this stage
within our setup, the total luminosities are dominated by
(anti-)neutrinos with energies above those for which opacities are
noticeably modified. It might partly be an artifact of the
approximations within the FMT neutrino treatment, where in particular
inelastic scattering reactions are neglected, which contribute to
redistributing neutrino energies and could thus lead to having more low
energy neutrinos in the spectrum. We also expect stronger modifications
of luminosities at later times, when emitted neutrinos have a mean
energy of the order of a few \si{\mev}.

\begin{figure*}
  \subfloat[Electron neutrino $\nu_e$, energy $E_\nu=\SI{2}{\mev}$]{\includegraphics[width=\columnwidth]{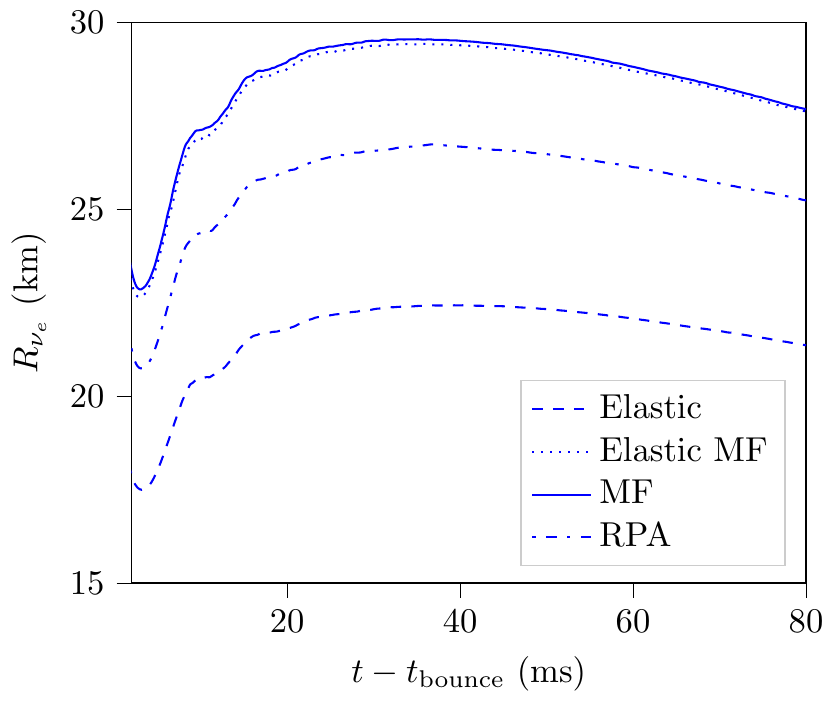}}
  \subfloat[Electron antineutrino $\bar\nu_e$, energy $E_\nu=\SI{2}{\mev}$]{
    \includegraphics[width=\columnwidth]{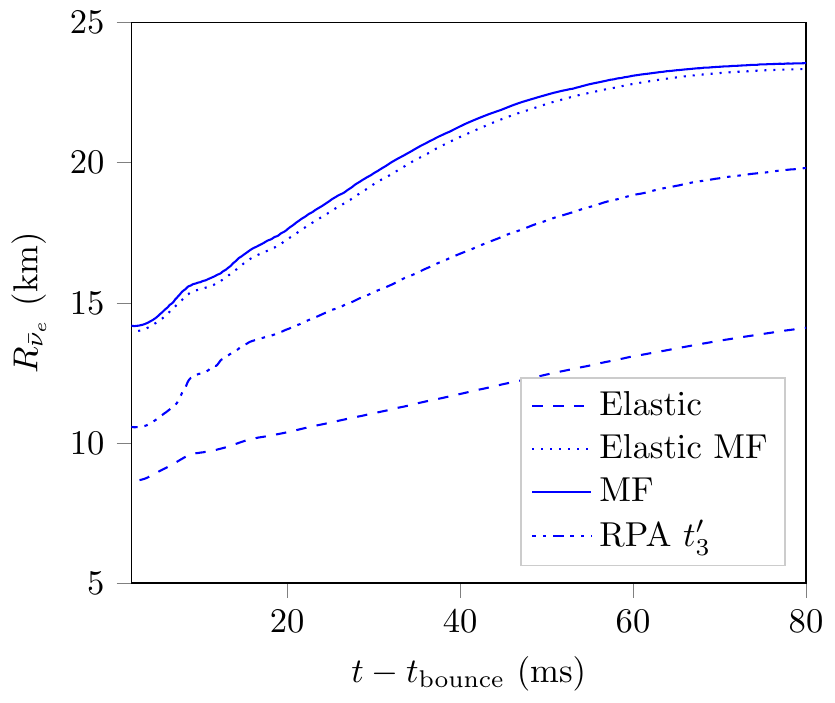}
  }\\
  \subfloat[Electron neutrino $\nu_e$, energy $E_\nu=\SI{14}{\mev}$]{
    \includegraphics[width=\columnwidth]{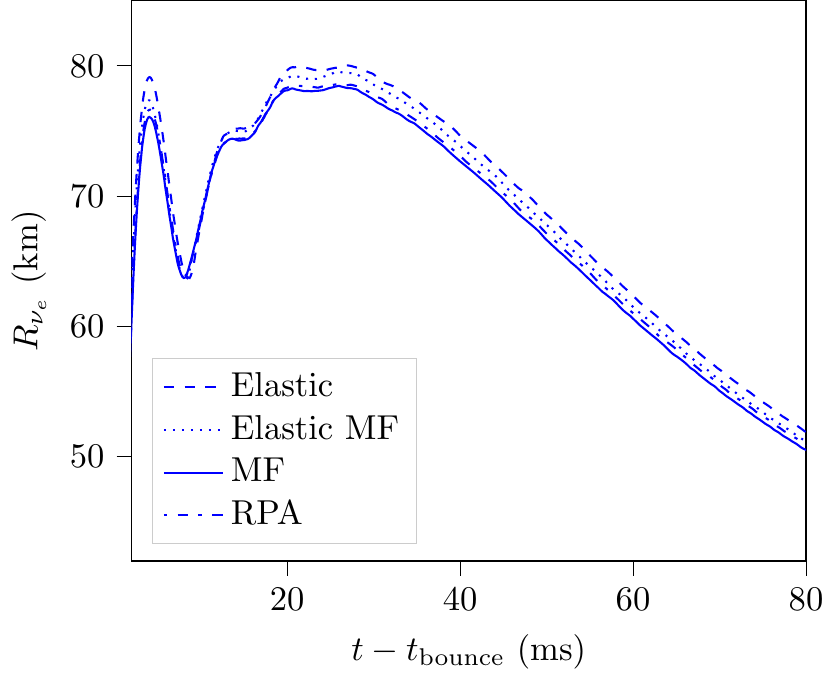}}
    \subfloat[Electron antineutrino $\bar\nu_e$, and energy $E_\nu=\SI{14}{\mev}$]{
    \includegraphics[width=\columnwidth]{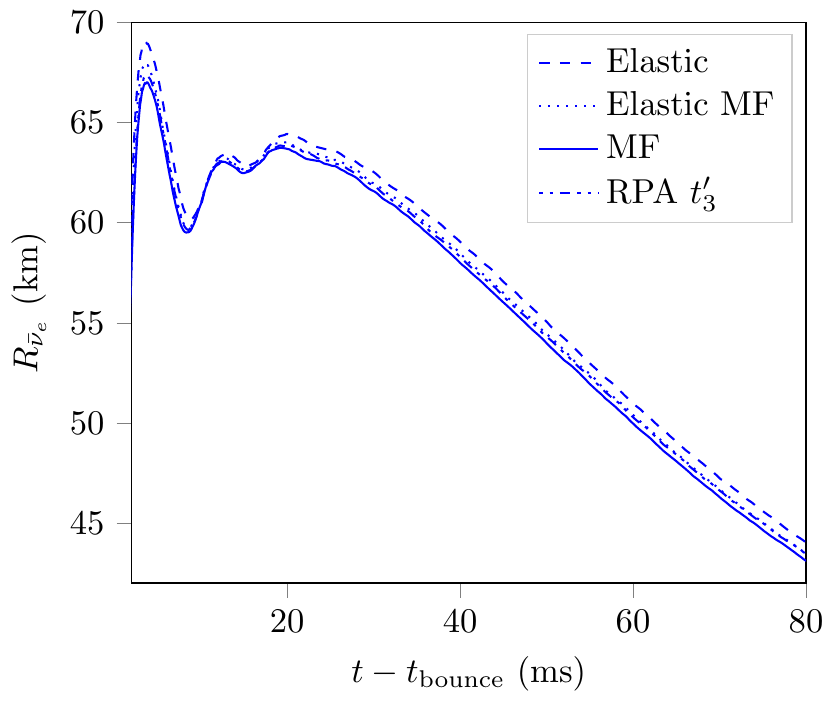} }
  \caption{Neutrinospheres radius  as a function of
    time after bounce, for two different neutrinos energies.}
  \label{fig:neutrinosph}
\end{figure*}

The modified opacities for low energy (anti-)neutrinos clearly affect
the location of the neutrinospheres. Here, we have defined them as the
radius where optical depth reaches $2/3$. In
Fig.~\ref{fig:neutrinosph}, upper panel, the neutrinospheres for
$\nu_e$ (left) and $\bar{\nu_e}$ (right) with an energy of $E_\nu = 2$
MeV are displayed, employing the different prescriptions for the
charged-current opacities. The results for the opacities for typical
conditions discussed in Section~\ref{ss:opacities} are clearly
reflected in the position of the neutrinosphere. Increased opacities
due to mean field effects compared with the basic elastic
approximation make it more difficult for (anti-)neutrinos to escape
and lead therefore to a neutrinosphere at larger radii. On the other
hand, a reduction of opacities within RPA with respect to mean field
facilitates escape and shifts the neutrinosphere again to smaller
radii. As anticipated, for larger neutrino energies, the difference
becomes smaller, see the example for $E_\nu = 14$ MeV in the lower
panel of Fig.~\ref{fig:neutrinosph}.

\section{Summary and discussion}\label{s:discussion}

Within this work we have computed opacities for $\nu_e$ and
$\bar{\nu}_e$ from charged current neutrino-nucleon interactions,
going beyond the elastic approximation and including nuclear
correlations in RPA. Please note that we do not pretend here that our
opacity data represent the ultimate description of nuclear
correlations. The differences in opacities induced by the different
prescriptions for evaluating the axial channel give an idea about the
uncertainties within RPA. Moreover, it only takes into account a
certain class of correlations, the long-range linear response. It is,
however, known since many years that these more accurate calculations
beyond the elastic approximation induce important changes in the
opacities in dense matter~\cite{Reddy:1997yr, Reddy:1998hb,
  Burrows:1998ek} and are therefore susceptible to modify the dynamics
of CCSN and matter composition in BNS mergers. Hence, we have
presented here a first step, proposing a scheme which allows to
incorporate accurate opacities directly into numerical simulations
which otherwise would be too time consuming to be calculated ``on
the fly''. We have been able to perform CCSN
simulations with consistently computed accurate charged current
(electron) neutrino nucleon interactions. We find noticeable
differences in the location of the neutrinospheres of low-energy
(anti-)neutrinos in the early post-bounce phase. In
\textcite{Fischer:2018kdt}, where rates in mean field with full phase
space are compared with the elastic approximation, during the longer
term proto-neutron star evolution in particular changes in the mean
energies of $\bar{\nu_e}$ and in the composition of the
neutrino-driven wind, i.e. the conditions for nucleosynthesis, are
pointed out. We expect the RPA correlations to impact these quantities
as well, a detailed study will be carried out in future work. It should be
stressed at this point that the modifications in the opacities with
respect to the commonly employed elastic approximation strongly depend
on neutrino energy and therefore cannot be cast into a grey correction
factor to the analytic expressions as previously implemented in
simulations.

After SN~1987A, much progress has been made and a possible neutrino
signal observed from a galactic supernova in present day detectors
would bear essential information about the core collapse mechanism and
neutrino properties~\cite{Mirizzi:2015eza}. It is therefore crucial
that models use accurate neutrino matter interaction rates. As
prospected since more than twenty years~\cite{Reddy:1997yr,
  Reddy:1998hb, Burrows:1998cg, Burrows:1998ek}, this work represents
a first step in enabling the use of these accurate rates in CCSN
simulations. It focuses on charged current neutrino nucleon opacities
and clearly shows that, indeed, commonly employed approximations break
down in the dense central part. These results encourage on the one
hand to go beyond our simplified setup for the simulations (spherical
symmetry, simplified neutrino transport) and to include these accurate
rates in more sophisticated simulations, in order to study in details
not only bounce properties but shock and post-bounce dynamics, too,
and the corresponding neutrino signal. To that end, the tables with
our opacity data are publicly available within the \textsc{Compose}
data base. A non-optimized implementation of the tables within our
code leads to maximally 50\% increase in computing time with respect
to standard analytic formulae. This is probably an upper limit since
the FMT does not include inelastic $\nu_e$-$e^\pm$ scattering which is
in general much more time consuming than charged-current opacities.
On the other hand, efforts should be pursued to extend the present
scheme for neutrino rates to other channels and offer ultimately the
possibility to include neutrino-matter interactions with state-of-the
art input physics within simulations. In particular, we have only
considered improved opacities for charged current neutrino-nucleon
interactions. Their increase in the dense regions makes them largely
dominant, whereas with the standard approximations neutral current
opacities are of the same order of magnitude. The sensitivity of CCSN
dynamics to the detailed treatment of the neutral currents has been
shown in \textcite{Melson:2015spa}, too, where a small change in the
coupling constants due to the strangeness content of the nucleon
decides upon explosion. A consistent implementation of improved
neutral current opacities~\cite{Reddy:1997yr, Reddy:1998hb,
  Burrows:1998cg, Navarro:1999jn, Margueron:2004sr, Horowitz:2006pj,
  Horowitz:2016gul} is thus in order and shall be investigated in
future work. Among others, we expect it to influence the relation
between neutrino trapping and the onset of $\beta$-equilibrium.

\begin{appendix}
  \section{Format of the opacity tables}
  \label{app:tables}
Data for (anti-)neutrino opacities are provided using the \hdf
data format \cite{hdf5}. Two \hdf groups, \texttt{nu} and
\texttt{nu\_bar}, contain the necessary data to evaluate the
respective opacities for $\nu_e$ and $\bar{\nu}_e$ via the
interpolation scheme discussed in Sec.~\ref{ss:interpolation}. In each
group are stored the following attributes: the information on the maximum number
$n_d^{\mathit{max}}$ ( attribute \texttt{nd\_max}) of domains in
$E_\nu$, the number of grid points for temperature $i_T$
(\texttt{pts\_t}), for baryon number density $j_B$ (\texttt{pts\_nb}) and
for electron fraction $k_Y$ (\texttt{pts\_ye}) as well as the number $n_p$
(\texttt{npts}) of interpolation coefficients. Within each
group, four datasets exist: \texttt{nd\_tny}, \texttt{enumin},
\texttt{enumax}, and \texttt{coeffs}. \texttt{nd\_tny} contains an
array of size ($i_T,j_B,k_Y$) with the number $n_d$ of domains in
$E_\nu$ for each point on the grid in $(T, n_B, Y_e)$ of the EoS
table. The two datasets \texttt{enumin} and \texttt{enumax} contain
arrays of size ($n_d^{\mathit{max}},i_T,j_B,k_Y$). The first $n_d$
nonzero entries contain, respectively, the logarithm of the minimum
and maximum energy in each domain for each point on the EoS grid. The
dataset \texttt{coeffs} is the array containing the coefficients $c_n$
in Eq.~(\ref{eq:interpolation}) in all domains and for the entire EoS
grid. The structure of the \hdf ~file is summarized in
Table~\ref{tab:coeffs}.

\begin{table}[t]
  \squeezetable
  \begin{tabular}{|c|c|c|}
    \hline  Name & Quantity & Type \\
     & & (size) \\ \hline\hline
    \texttt{npts} & $n_p=N+1$: number of   & integer \\  &  interpolation coefficients $c_n$ & \\  &  with $n \in (0,\dots, N)$ & \\ \hline
    \texttt{pts\_t} & number $i_T$ of points in $T$ & integer \\ \hline
    \texttt{pts\_nb} & number $j_B$ of points in $n_B$ & integer \\ \hline
    \texttt{pts\_ye} & number $k_Y$ of points in $Y_e$  &integer \\ \hline
    \texttt{nd\_tny} & number $n_d$ of domains & array of integers \\ &  in $E_\nu$  for each &  ($i_T,j_B,k_Y$) \\
     & grid point in $(T,n_B,Y_e)$ &  \\ \hline
    \texttt{nd\_max} & $n_{d}^{\mathit{max}} = \mathrm{max}(n_d)$ :
                       maximum & array of integers \\
    &  number of domains in $E_\nu$ & \\ \hline
    \texttt{enumin} & logarithm of minimum $E_\nu$  & array of doubles \\
     & in each domain for each & ($n_d^{\mathrm{max}},i_T,j_B,k_Y$) \\
     & grid point in $(T,n_B,Y_e)$ &  \\ \hline
    \texttt{enumax} & logarithm of maximum $E_\nu$ & array of doubles \\
    & in each domain for each &  ($n_{d}^{\mathrm{max}},i_T,j_B,k_Y$) \\
     & grid point in $(T,n_B,Y_e)$ &  \\ \hline
    \texttt{coeffs} & interpolation coefficients $c_n$ & array of doubles \\ &  in each domain and for each & ($n_p,n_{d}^{\mathrm{max}},i_T,j_B,k_Y$) \\
     & grid point in $(T,n_B,Y_e)$ &  \\ \hline
  \end{tabular}
  \caption{Summary of the structure of the \hdf~file storing the
    opacity data for $\nu_e$ (group \texttt{nu}) and $\bar{\nu}_e$
    (group \texttt{nu\_bar}). \label{tab:coeffs}}
\end{table}
We have considered that the detail values of the opacity were
irrelevant when the mean free path became (much) larger than the
extension of the studied astrophysical object, and therefore a lower bound for
$\kappa$ has been introduced: for all
$\kappa < \kappa_{\mathit{limit}} = 10^{-10}$/km, we have set
$\kappa = \kappa_{\mathit{limit}}$ in the data tables.

  \section{Lindhard function in asymmetric matter}
  \label{app:lindhard}
Eq.~(\ref{eq:lindhard}) for the Lindhard function can be further developed by substituting in the last term $\vec{k} \to - \vec{k} - \vec{q}$
\begin{align}
L(q) &= \lim_{\eta\to 0} \int \frac{d^3 k}{(2 \pi)^3}\left(\frac{f_F(\ep-\mu_p^*)}{\tilde{q_0} + i \eta + \ep - \en}\right.\nonumber\\ & \qquad\qquad\qquad{} - \left.
\frac{f_F(\epsilon_k^n-\mu_n^*)}{\tilde{q_0} + i \eta + \epsilon_{k+q}^p - \epsilon_k^n}\right)
~.
\end{align}
The angular integration can be performed analytically using $\epsilon_{k+q}^i = (\vec{k}+\vec{q})^2/(2 m_i^*) + m_i^*$. Thus,
\begin{align}
  L(q) &= \frac{1}{4 \pi^2|\vec{q}|} \; \lim_{\eta \to 0}\int k dk \times  \nonumber \\ &
  \qquad \qquad{}\left( \int_{\epsilon_-^n}^{\epsilon_+^n} d x \;
  m_n \frac{f_F(\ep - \mu^*_p)}{\tilde{q_0} + i \eta  + \ep - x}\right. \nonumber \\ & \qquad\qquad{} -
  \int_{\epsilon_-^p}^{\epsilon_+^p} d x
\left. m_p \frac{f_F(\epsilon_k^n-\mu^*_n)}{\tilde{q_0} + i \eta  + x - \epsilon_k^n}\right)\nonumber \\
&= -\frac{1}{4 \pi^2|\vec{q}|}\;\lim_{\eta \to 0} \int k dk \times\nonumber \\ & \left(m_n f_F(\ep - \mu_p^*) \log\left(\frac{\epsilon_+^n - \tilde{q_0}- i\eta  - \ep}{\epsilon_-^n - \tilde{q_0} - i \eta - \ep}\right) \right.\nonumber \\ &  +\left. m_p f_F(\epsilon_k^n-\mu_n^*) \log\left(\frac{\epsilon_+^p + \tilde{q_0} + i \eta - \epsilon_k^n}{\epsilon_-^p + \tilde{q_0} + i \eta - \epsilon_k^n}\right)\right)~.
\label{eq:lindhard2}
\end{align}
The integration boundaries are given by
\begin{equation}
\epsilon_{\pm}^i = \frac{(|\vec{k}| \pm |\vec{q}|)^2}{2 m_i^*} + m^*_i~.
\end{equation}
For the real part, the remaining integration over momentum in
Eq.~(\ref{eq:lindhard2}) has to be carried out
numerically. Alternatively, the real part can be obtained from a
dispersion integral with the analytic expression of the imaginary
part, see Eq.~(\ref{eq:imlim}) below.

Using the technique indicated in \textcite{Reddy:1997yr} we can give
an analytic expression for the imaginary part of the Lindhard function
in the case of non-interacting particles or in mean field. To that end
we rewrite it as, see Eq.~(\ref{eq:lindhard2})
\begin{align}
  \mathrm{Im}\, L(q) &= -\frac{1}{4 \pi |\vec{q}|} \int k dk \times \nonumber \\ &
  \left(\int_{\epsilon_-^n}^{\epsilon_+^n} d x \; m_n f_F(\ep - \mu^*_p)\, \delta(\tilde{q_0} + \ep - x) \right. \nonumber \\ & - \int_{\epsilon_-^p}^{\epsilon_+^p} d x
\left. m_p f_F(\epsilon_k^n-\mu^*_n)\, \delta(\tilde{q_0} + x - \epsilon_k^n)\right)
\end{align}
The angular integration can be evaluated with the help of the
$\delta$-function and we obtain
\begin{align}
\mathrm{Im} L(q) &= -\frac{1}{4 \pi \, |\vec{q}|} \left(m^*_n \int_{k_-^p}^{k_+^p} k dk\, \,f_F(\ep - \mu^*_p)\right.\nonumber\\ & \qquad{} \left. - m^*_p \int_{k_-^n}^{k_+^n} k dk \, f_F (\enk - \mu^*_n)\right)~,
\end{align}
where $k_{\pm}^i$ is given by the boundaries on the angular integration,
\begin{align}
k_\pm^i &= \left|\frac{ \pm m_i^* |\vec{q}| + X }{m_n^* - m_p^*}\right| \\ X &= \left(m_n^* m_p^*\, (\vec{q}^2 + 2\,(m_n^* - m_p^*) \, (m_n^* - m_p^* - \tilde{q_0}))\right)^{1/2}\nonumber
\end{align}
 The remaining integration can be carried by employing Eq. (20) of
 Ref.~\cite{Reddy:1997yr} with $ x_i T = \vec{k}^2/(2 m_i^*) + m_i^* -
 \mu_i^*$
\begin{equation}
  \mathrm{Im} L(q) = -\frac{T m_n^* m_p^*}{4 \pi\,|\vec{q}|} \left(\int_{x_-^p}^{x_+^p} \frac{dx}{e^x + 1} - \int_{x_-^n}^{x_+^n} \frac{dx}{e^x + 1} \right)~,
\end{equation}
where $x_\pm^i = ((k^i_\pm)^2/(2 m_i^*) + m_i^* - \mu_i^*)/T$. The final integration leads to
\begin{equation}
  \mathrm{Im} L(q) = \frac{T m_n^* m_p^*}{4 \pi\,|\vec{q}|} \left\{\ln\left(\frac{1 + e^{-x_+^n}}{ 1+ e^{-x_-^n}}\right) -
 \ln\left(\frac{1 + e^{-x_+^p}}{ 1+ e^{-x_-^p}}  \right)\right\}~.
\label{eq:imlim}
\end{equation}
The result, apart from being written in a more symmetric way, agrees
with that from \textcite{Reddy:1997yr}, Eq. (41), if the different
definition of $q_0$ is considered, which cancels here the explicit
$\mu_p - \mu_n$-term in the argument of the logarithm. In addition,
the factor $(1 + f_B(q_0))$ is absent since it is explicitly accounted
for in Eq.~(\ref{eq:dfnudt}).

\section{Opacity expressions}
\label{app:opacities}

As mentioned in section~\ref{ss:formalism}, the contribution of positronic
processes to the opacities can be obtained straight\-forwardly
by replacing $(E_e,\vec{k_e}) \to - (E_e,\vec{k_e})$ in the
expressions for electronic processes. The complete emissivity and mean
free path for neutrinos including reactions $p + e^- \leftrightarrow n
+ \nu_e$ and $ p \leftrightarrow n + e^+ + \nu_e$ becomes then
  \begin{align}
    j(E_\nu) =& - \frac{G_F^2 V_{ud}^2}{8} \int \frac{d^3 k_e}{ (2 \pi)^3} \frac{1}{E_e E_\nu} \times \nonumber \\
 & \left\{ L^{\lambda\sigma} \mathrm{Im} \Pi^{R}_{\lambda\sigma}(q)
    \times \right. \nonumber \\ & \qquad f_F (E_e - \mu_e))\,( 1+ f_B(q_{0})) \nonumber \\ +& L^{\lambda\sigma}\mathrm{Im} \Pi^{R}_{\lambda\sigma}(q^{+}) \times \nonumber \\ & \left. \qquad (1- f_F(E_e + \mu_e)) \,(1 + f_B (q_{0}^{+})) \right\}
    \nonumber \\
    \frac{1}{\lambda(E_\nu)} =& - \frac{G_F^2 V_{ud}^2}{8} \int \frac{d^3 k_e}{ (2 \pi)^3} \frac{1}{E_e E_\nu} \times \nonumber \\ &
\left\{ L^{\lambda\sigma} \mathrm{Im} \Pi^{R}_{\lambda\sigma}(q)
    \times \right. \nonumber \\ & \qquad(1- f_F (E_e - \mu_e))\, f_B(q_{0}) \nonumber \\ +& L^{\lambda\sigma}\mathrm{Im} \Pi^{R}_{\lambda\sigma}(q^{+}) \times \nonumber \\ & \left. \qquad f_F(E_e + \mu_e) \, f_B (q_{0}^{+}) \right\}
    \label{eq:nuemissivity_complete}
\end{align}
with $q^{+} = (-E_e - E_\nu - \mu_e+ \mu_\nu,- \vec{k_e} -
\vec{k_\nu})$. For anti-neutrinos it becomes, including $n
\leftrightarrow p + e^- + \bar{\nu}_e$ and $ n + e^+ \leftrightarrow p
+ \bar{\nu}_e$ reactions
\begin{align}
  \bar{\jmath}(E_\nu) =& - \frac{G_F^2 V_{ud}^2}{8} \int \frac{d^3 k_e}{ (2 \pi)^3} \frac{1}{E_e E_\nu} \times \nonumber \\ & \left\{ L^{\lambda\sigma} \mathrm{Im} \Pi^{R}_{\lambda\sigma}(\bar{q}) \times
  \right.\nonumber \\ &\qquad (1- f_F(E_e - \mu_e)) \, f_B (\bar{q_0}) \nonumber \\
  + & L^{\lambda\sigma} \mathrm{Im} \Pi^{R}_{\lambda\sigma}(\bar{q}^+) \times \nonumber \\ & \qquad \left. f_F(E_e + \mu_e)\, f_B (\bar{q_0}^+)
  \right\} \nonumber \\
  \frac{1}{\bar{\lambda}(E_\nu)} =& - \frac{G_F^2 V_{ud}^2}{8} \int \frac{d^3 k_e}{ (2 \pi)^3} \frac{1}{E_e E_\nu} \times \nonumber \\ & \left\{ L^{\lambda\sigma} \mathrm{Im} \Pi^{R}_{\lambda\sigma}(\bar{q}) \times
  \right. \nonumber \\ &\qquad  f_F (E_e - \mu_e)\, (1+ f_B(\bar{q_0}))
  \nonumber \\ + & L^{\lambda\sigma} \mathrm{Im} \Pi^{R}_{\lambda\sigma}(\bar{q}^+) \times \nonumber \\ & \qquad (1-f_F (E_e + \mu_e))\, (1+ f_B(\bar{q_0}^{+}))
\label{eq:nubaremissivity_complete}
\end{align}
with $\bar{q}^+ = (-E_e + E_\nu - \mu_e+ \mu_\nu,- \vec{k_e} + \vec{k_\nu})$.
The simple properties, cf Eq.~(\ref{eq:detailedbalance}),
\begin{align}
  & (1-f_F(E_e+\mu_e))\, (1+f_B(q_0^+)) =\nonumber \\  &\quad f_B(q_0^+) f_F(E_e+\mu_e) \exp((- E_\nu + \mu_\nu)/T)~,\nonumber \\
  & (1-f_F(E_e+\mu_e))\, (1+f_B(\bar{q_0}^+)) =\nonumber \\  &\quad f_B(\bar{q_0}^+) f_F(E_e+\mu_e) \exp(( E_\nu + \mu_\nu)/T)~,
  \label{eq:detailedbalance_bis}
\end{align}
reflect detailed balance for positronic processes as it should. In terms of emissivity and mean free path detailed balance reads
\begin{align}
  j(E_\nu)& = \frac{\exp((-E_\nu + \mu_\nu)/T)}{\lambda(E_\nu)} \nonumber \\
  \bar{\jmath}(E_\nu) &= \frac{\exp((-E_\nu - \mu_\nu)/T)}{\bar{\lambda}(E_\nu)} ~.
  \label{eq:db}
\end{align}
For practical purposes, the integration over
electron momenta can be transformed into an integration over $q_0$ and
$|\vec{q}|$. For instance, considering the reactions with electrons and neutrinos, $p+ e^- \leftrightarrow n + \nu_e$, we can use
\begin{equation}
\vec{q}^2 = \vec{k_e}^2 + E_\nu^2 - 2 |\vec{k_e}| E_\nu z~,
\end{equation}
and $q_0 = E_e - E_\nu - \mu_e + \mu_\nu$ to obtain
\begin{align}
\int \frac{d^3 k_e}{(2 \pi)^3} &= \frac{1}{4 \pi^2} \int |\vec{k_e}| E_e d E_e \int_{-1}^1 dz \nonumber \\ &= \frac{1}{4 \pi^2} \int_{m_e - E_\nu - \mu_e + \mu_\nu}^\infty \frac{E_e}{E_\nu} dq_0 \int_{|k-E_\nu|}^{k+E_\nu} |\vec{q}| d|\vec{q}|~,
\end{align}
with $k = |\vec{k_e}|$.

In elastic approximation, cf Eq.~(\ref{eq:dfnudtelastic}), neutrino emissivity becomes
\begin{align}
j(E_\nu) &=  \frac{G_F^2 V_{ud}^2}{\pi} \, (g_V^2 + 3 g_A^2) (n_p - n_n)\, (1+f_B(q_0)) \times \nonumber \\ & \left( k_{e^-} E_{e^-}\, f_F(E_{e^-}-\mu_e) \right. \nonumber \\  & + \left.
 k_{e^+} E_{e^+}\,
  (1-f_F(E_{e^+} + \mu_e)) \right)~,
\label{eq:jelasticnu}
\end{align}
with $E_{e^\pm} = \mp (E_\nu + m_n - m_p)$ and $q_0 = m_n - m_p + \mu_p - \mu_n$. $k_{e^\pm} = \sqrt{E_{e^\pm}^2 - m_e^2}$ denotes the momentum of the charged lepton.
For anti-neutrinos we have
\begin{align}
\bar{\jmath}(E_\nu) &=  \frac{G_F^2 V_{ud}^2}{\pi} \, (g_V^2 + 3 g_A^2) (n_p - n_n) \, f_B(q_0)\times \nonumber \\ & \left( k_{e^-} E_{e^-}\, (1-f_F(E_{e^-}-\mu_e)) \right. \nonumber \\ & + \left.
k_{e^+} E_{e^+}\,
f_F(E_{e^+} + \mu_e) \right)~,
\label{eq:jelasticnubar}
\end{align}
with $E_{e^\pm} = \mp(-E_\nu + m_n - m_p)$ and
$q_0 = m_n - m_p + \mu_p - \mu_n$. The mean free paths are then given by
Eqs.~(\ref{eq:db}). Corrected for mean field effects this becomes for neutrinos
\begin{align}
j(E_\nu) &=  \frac{G_F^2 V_{ud}^2}{\pi} \, (g_V^2 + 3 g_A^2) \eta_{pn} \times \nonumber \\ & \left( k_{e^-} E_{e^-} \, f_F(E_{e^-}-\mu_e) \right. \nonumber \\   & +  \left. k_{e^+} E_{e^+}\,
  (1-f_F(E_{e^+} + \mu_e)) \right)~,
\label{eq:jelasticnumf}
\end{align}
with $E_{e^\pm} = \mp (E_\nu + m_n^* - m_p^* + U_n - U_p)$ and $k_{e^\pm} = \sqrt{E_{e^\pm}^2 - m_e^2}$.
For anti-neutrinos
\begin{align}
\bar{\jmath}(E_\nu) &=  \frac{G_F^2 V_{ud}^2}{\pi} \, (g_V^2 + 3 g_A^2) \eta_{np}\times \nonumber \\ & \left( k_{e^-}E_{e^-}\, (1-f_F(E_{e^-}-\mu_e)) \right. \nonumber \\ +& \left. k_{e^+} E_{e^+}\,
f_F(E_{e^+} + \mu_e) \right)~,
\label{eq:jelasticnubarmf}
\end{align}
with $E_{e^\pm} = \mp(-E_\nu + m_n^* - m_p^* + U_n - U_p)$.

\end{appendix}

\begin{acknowledgments}
  We would like to thank H.-T. Janka for encouraging this work,
  R. Bollig, H.-T. Janka and M. Urban enlightening discussions
  and B. Müller for providing us with the FMT code. The research
  leading to these results has received funding from the PICS07889; it
  was also partially supported by the PHAROS European Science and
  Technology (COST) Action CA16214 and the Observatoire de Paris
  through the action fédératrice ``PhyFog''. This work was granted
  access to the computing resources of MesoPSL financed by the Region
  Ile de France and the project Equip@Meso (reference
  ANR-10-EQPX-29-01) of the programme Investissements d’Avenir
  supervised by the Agence Nationale pour la Recherche.
\end{acknowledgments}

\bibliography{biblio}

\begin{thebibliography}{70}%
\makeatletter
\providecommand \@ifxundefined [1]{%
 \@ifx{#1\undefined}
}%
\providecommand \@ifnum [1]{%
 \ifnum #1\expandafter \@firstoftwo
 \else \expandafter \@secondoftwo
 \fi
}%
\providecommand \@ifx [1]{%
 \ifx #1\expandafter \@firstoftwo
 \else \expandafter \@secondoftwo
 \fi
}%
\providecommand \natexlab [1]{#1}%
\providecommand \enquote  [1]{``#1''}%
\providecommand \bibnamefont  [1]{#1}%
\providecommand \bibfnamefont [1]{#1}%
\providecommand \citenamefont [1]{#1}%
\providecommand \href@noop [0]{\@secondoftwo}%
\providecommand \href [0]{\begingroup \@sanitize@url \@href}%
\providecommand \@href[1]{\@@startlink{#1}\@@href}%
\providecommand \@@href[1]{\endgroup#1\@@endlink}%
\providecommand \@sanitize@url [0]{\catcode `\\12\catcode `\$12\catcode
  `\&12\catcode `\#12\catcode `\^12\catcode `\_12\catcode `\%12\relax}%
\providecommand \@@startlink[1]{}%
\providecommand \@@endlink[0]{}%
\providecommand \url  [0]{\begingroup\@sanitize@url \@url }%
\providecommand \@url [1]{\endgroup\@href {#1}{\urlprefix }}%
\providecommand \urlprefix  [0]{URL }%
\providecommand \Eprint [0]{\href }%
\providecommand \doibase [0]{http://dx.doi.org/}%
\providecommand \selectlanguage [0]{\@gobble}%
\providecommand \bibinfo  [0]{\@secondoftwo}%
\providecommand \bibfield  [0]{\@secondoftwo}%
\providecommand \translation [1]{[#1]}%
\providecommand \BibitemOpen [0]{}%
\providecommand \bibitemStop [0]{}%
\providecommand \bibitemNoStop [0]{.\EOS\space}%
\providecommand \EOS [0]{\spacefactor3000\relax}%
\providecommand \BibitemShut  [1]{\csname bibitem#1\endcsname}%
\let\auto@bib@innerbib\@empty
\bibitem [{\citenamefont {Abbott}\ \emph {et~al.}(2017)\citenamefont {Abbott}
  \emph {et~al.}}]{abbott_17}%
  \BibitemOpen
  \bibfield  {author} {\bibinfo {author} {\bibfnamefont {B.}~\bibnamefont
  {Abbott}} \emph {et~al.} (\bibinfo {collaboration} {Virgo, LIGO
  Scientific}),\ }\href {\doibase 10.1103/PhysRevLett.119.161101} {\bibfield
  {journal} {\bibinfo  {journal} {Phys. Rev. Lett.}\ }\textbf {\bibinfo
  {volume} {119}},\ \bibinfo {pages} {161101} (\bibinfo {year}
  {2017})}\BibitemShut {NoStop}%
\bibitem [{\citenamefont {Bruenn}(1985)}]{Bruenn:1985en}%
  \BibitemOpen
  \bibfield  {author} {\bibinfo {author} {\bibfnamefont {S.~W.}\ \bibnamefont
  {Bruenn}},\ }\href {\doibase 10.1086/191056} {\bibfield  {journal} {\bibinfo
  {journal} {Astrophys. J. Suppl.}\ }\textbf {\bibinfo {volume} {58}},\
  \bibinfo {pages} {771} (\bibinfo {year} {1985})}\BibitemShut {NoStop}%
\bibitem [{\citenamefont {Rosswog}\ and\ \citenamefont
  {Liebendoerfer}(2003)}]{Rosswog:2003rv}%
  \BibitemOpen
  \bibfield  {author} {\bibinfo {author} {\bibfnamefont {S.}~\bibnamefont
  {Rosswog}}\ and\ \bibinfo {author} {\bibfnamefont {M.}~\bibnamefont
  {Liebendoerfer}},\ }\href {\doibase 10.1046/j.1365-8711.2003.06579.x}
  {\bibfield  {journal} {\bibinfo  {journal} {Mon. Not. Roy. Astron. Soc.}\
  }\textbf {\bibinfo {volume} {342}},\ \bibinfo {pages} {673} (\bibinfo {year}
  {2003})}\BibitemShut {NoStop}%
\bibitem [{\citenamefont {Burrows}\ \emph {et~al.}(2006)\citenamefont
  {Burrows}, \citenamefont {Reddy},\ and\ \citenamefont
  {Thompson}}]{Burrows:2004vq}%
  \BibitemOpen
  \bibfield  {author} {\bibinfo {author} {\bibfnamefont {A.}~\bibnamefont
  {Burrows}}, \bibinfo {author} {\bibfnamefont {S.}~\bibnamefont {Reddy}}, \
  and\ \bibinfo {author} {\bibfnamefont {T.~A.}\ \bibnamefont {Thompson}},\
  }\href {\doibase 10.1016/j.nuclphysa.2004.06.012} {\bibfield  {journal}
  {\bibinfo  {journal} {Nucl. Phys.}\ }\textbf {\bibinfo {volume} {A777}},\
  \bibinfo {pages} {356} (\bibinfo {year} {2006})}\BibitemShut {NoStop}%
\bibitem [{\citenamefont {Schmitt}\ and\ \citenamefont
  {Shternin}(2018)}]{Schmitt:2017efp}%
  \BibitemOpen
  \bibfield  {author} {\bibinfo {author} {\bibfnamefont {A.}~\bibnamefont
  {Schmitt}}\ and\ \bibinfo {author} {\bibfnamefont {P.}~\bibnamefont
  {Shternin}},\ }\href {\doibase 10.1007/978-3-319-97616-7_9} {\bibfield
  {journal} {\bibinfo  {journal} {Astrophys. Space Sci. Libr.}\ }\textbf
  {\bibinfo {volume} {457}},\ \bibinfo {pages} {455} (\bibinfo {year}
  {2018})}\BibitemShut {NoStop}%
\bibitem [{\citenamefont {Horowitz}(2002)}]{Horowitz:2001xf}%
  \BibitemOpen
  \bibfield  {author} {\bibinfo {author} {\bibfnamefont {C.~J.}\ \bibnamefont
  {Horowitz}},\ }\href {\doibase 10.1103/PhysRevD.65.043001} {\bibfield
  {journal} {\bibinfo  {journal} {Phys. Rev.}\ }\textbf {\bibinfo {volume}
  {D65}},\ \bibinfo {pages} {043001} (\bibinfo {year} {2002})}\BibitemShut
  {NoStop}%
\bibitem [{\citenamefont {Horowitz}(1997)}]{Horowitz:1996ci}%
  \BibitemOpen
  \bibfield  {author} {\bibinfo {author} {\bibfnamefont {C.~J.}\ \bibnamefont
  {Horowitz}},\ }\href {\doibase 10.1103/PhysRevD.55.4577} {\bibfield
  {journal} {\bibinfo  {journal} {Phys. Rev.}\ }\textbf {\bibinfo {volume}
  {D55}},\ \bibinfo {pages} {4577} (\bibinfo {year} {1997})}\BibitemShut
  {NoStop}%
\bibitem [{\citenamefont {Bruenn}\ and\ \citenamefont
  {Mezzacappa}(1997)}]{Bruenn:1997jv}%
  \BibitemOpen
  \bibfield  {author} {\bibinfo {author} {\bibfnamefont {S.~W.}\ \bibnamefont
  {Bruenn}}\ and\ \bibinfo {author} {\bibfnamefont {A.}~\bibnamefont
  {Mezzacappa}},\ }\href {\doibase 10.1103/PhysRevD.56.7529} {\bibfield
  {journal} {\bibinfo  {journal} {Phys. Rev.}\ }\textbf {\bibinfo {volume}
  {D56}},\ \bibinfo {pages} {7529} (\bibinfo {year} {1997})}\BibitemShut
  {NoStop}%
\bibitem [{\citenamefont {Reddy}\ \emph {et~al.}(1998)\citenamefont {Reddy},
  \citenamefont {Prakash},\ and\ \citenamefont {Lattimer}}]{Reddy:1997yr}%
  \BibitemOpen
  \bibfield  {author} {\bibinfo {author} {\bibfnamefont {S.}~\bibnamefont
  {Reddy}}, \bibinfo {author} {\bibfnamefont {M.}~\bibnamefont {Prakash}}, \
  and\ \bibinfo {author} {\bibfnamefont {J.~M.}\ \bibnamefont {Lattimer}},\
  }\href {\doibase 10.1103/PhysRevD.58.013009} {\bibfield  {journal} {\bibinfo
  {journal} {Phys. Rev.}\ }\textbf {\bibinfo {volume} {D58}},\ \bibinfo {pages}
  {013009} (\bibinfo {year} {1998})}\BibitemShut {NoStop}%
\bibitem [{\citenamefont {Roberts}\ \emph {et~al.}(2012)\citenamefont
  {Roberts}, \citenamefont {Reddy},\ and\ \citenamefont
  {Shen}}]{Roberts:2012um}%
  \BibitemOpen
  \bibfield  {author} {\bibinfo {author} {\bibfnamefont {L.~F.}\ \bibnamefont
  {Roberts}}, \bibinfo {author} {\bibfnamefont {S.}~\bibnamefont {Reddy}}, \
  and\ \bibinfo {author} {\bibfnamefont {G.}~\bibnamefont {Shen}},\ }\href
  {\doibase 10.1103/PhysRevC.86.065803} {\bibfield  {journal} {\bibinfo
  {journal} {Phys. Rev.}\ }\textbf {\bibinfo {volume} {C86}},\ \bibinfo {pages}
  {065803} (\bibinfo {year} {2012})}\BibitemShut {NoStop}%
\bibitem [{\citenamefont {Martinez-Pinedo}\ \emph {et~al.}(2012)\citenamefont
  {Martinez-Pinedo}, \citenamefont {Fischer}, \citenamefont {Lohs},\ and\
  \citenamefont {Huther}}]{MartinezPinedo:2012rb}%
  \BibitemOpen
  \bibfield  {author} {\bibinfo {author} {\bibfnamefont {G.}~\bibnamefont
  {Martinez-Pinedo}}, \bibinfo {author} {\bibfnamefont {T.}~\bibnamefont
  {Fischer}}, \bibinfo {author} {\bibfnamefont {A.}~\bibnamefont {Lohs}}, \
  and\ \bibinfo {author} {\bibfnamefont {L.}~\bibnamefont {Huther}},\ }\href
  {\doibase 10.1103/PhysRevLett.109.251104} {\bibfield  {journal} {\bibinfo
  {journal} {Phys. Rev. Lett.}\ }\textbf {\bibinfo {volume} {109}},\ \bibinfo
  {pages} {251104} (\bibinfo {year} {2012})}\BibitemShut {NoStop}%
\bibitem [{\citenamefont {Hannestad}\ and\ \citenamefont
  {Raffelt}(1998)}]{Hannestad:1997gc}%
  \BibitemOpen
  \bibfield  {author} {\bibinfo {author} {\bibfnamefont {S.}~\bibnamefont
  {Hannestad}}\ and\ \bibinfo {author} {\bibfnamefont {G.}~\bibnamefont
  {Raffelt}},\ }\href {\doibase 10.1086/306303} {\bibfield  {journal} {\bibinfo
   {journal} {Astrophys. J.}\ }\textbf {\bibinfo {volume} {507}},\ \bibinfo
  {pages} {339} (\bibinfo {year} {1998})}\BibitemShut {NoStop}%
\bibitem [{\citenamefont {Fischer}\ \emph {et~al.}(2020)\citenamefont
  {Fischer}, \citenamefont {Guo}, \citenamefont {Dzhioev}, \citenamefont
  {Martínez-Pinedo}, \citenamefont {Wu}, \citenamefont {Lohs},\ and\
  \citenamefont {Qian}}]{Fischer:2018kdt}%
  \BibitemOpen
  \bibfield  {author} {\bibinfo {author} {\bibfnamefont {T.}~\bibnamefont
  {Fischer}}, \bibinfo {author} {\bibfnamefont {G.}~\bibnamefont {Guo}},
  \bibinfo {author} {\bibfnamefont {A.~A.}\ \bibnamefont {Dzhioev}}, \bibinfo
  {author} {\bibfnamefont {G.}~\bibnamefont {Martínez-Pinedo}}, \bibinfo
  {author} {\bibfnamefont {M.-R.}\ \bibnamefont {Wu}}, \bibinfo {author}
  {\bibfnamefont {A.}~\bibnamefont {Lohs}}, \ and\ \bibinfo {author}
  {\bibfnamefont {Y.-Z.}\ \bibnamefont {Qian}},\ }\href {\doibase
  10.1103/PhysRevC.101.025804} {\bibfield  {journal} {\bibinfo  {journal}
  {Phys. Rev.}\ }\textbf {\bibinfo {volume} {C101}},\ \bibinfo {pages} {025804}
  (\bibinfo {year} {2020})}\BibitemShut {NoStop}%
\bibitem [{\citenamefont {Yakovlev}\ \emph {et~al.}(2001)\citenamefont
  {Yakovlev}, \citenamefont {Kaminker}, \citenamefont {Gnedin},\ and\
  \citenamefont {Haensel}}]{Yakovlev:2000jp}%
  \BibitemOpen
  \bibfield  {author} {\bibinfo {author} {\bibfnamefont {D.~G.}\ \bibnamefont
  {Yakovlev}}, \bibinfo {author} {\bibfnamefont {A.~D.}\ \bibnamefont
  {Kaminker}}, \bibinfo {author} {\bibfnamefont {O.~Y.}\ \bibnamefont
  {Gnedin}}, \ and\ \bibinfo {author} {\bibfnamefont {P.}~\bibnamefont
  {Haensel}},\ }\href {\doibase 10.1016/S0370-1573(00)00131-9} {\bibfield
  {journal} {\bibinfo  {journal} {Phys. Rept.}\ }\textbf {\bibinfo {volume}
  {354}},\ \bibinfo {pages} {1} (\bibinfo {year} {2001})}\BibitemShut {NoStop}%
\bibitem [{\citenamefont {Burrows}\ and\ \citenamefont
  {Sawyer}(1998)}]{Burrows:1998cg}%
  \BibitemOpen
  \bibfield  {author} {\bibinfo {author} {\bibfnamefont {A.}~\bibnamefont
  {Burrows}}\ and\ \bibinfo {author} {\bibfnamefont {R.~F.}\ \bibnamefont
  {Sawyer}},\ }\href {\doibase 10.1103/PhysRevC.58.554} {\bibfield  {journal}
  {\bibinfo  {journal} {Phys. Rev.}\ }\textbf {\bibinfo {volume} {C58}},\
  \bibinfo {pages} {554} (\bibinfo {year} {1998})}\BibitemShut {NoStop}%
\bibitem [{\citenamefont {Burrows}\ and\ \citenamefont
  {Sawyer}(1999)}]{Burrows:1998ek}%
  \BibitemOpen
  \bibfield  {author} {\bibinfo {author} {\bibfnamefont {A.}~\bibnamefont
  {Burrows}}\ and\ \bibinfo {author} {\bibfnamefont {R.~F.}\ \bibnamefont
  {Sawyer}},\ }\href {\doibase 10.1103/PhysRevC.59.510} {\bibfield  {journal}
  {\bibinfo  {journal} {Phys. Rev.}\ }\textbf {\bibinfo {volume} {C59}},\
  \bibinfo {pages} {510} (\bibinfo {year} {1999})}\BibitemShut {NoStop}%
\bibitem [{\citenamefont {Reddy}\ \emph {et~al.}(1999)\citenamefont {Reddy},
  \citenamefont {Prakash}, \citenamefont {Lattimer},\ and\ \citenamefont
  {Pons}}]{Reddy:1998hb}%
  \BibitemOpen
  \bibfield  {author} {\bibinfo {author} {\bibfnamefont {S.}~\bibnamefont
  {Reddy}}, \bibinfo {author} {\bibfnamefont {M.}~\bibnamefont {Prakash}},
  \bibinfo {author} {\bibfnamefont {J.~M.}\ \bibnamefont {Lattimer}}, \ and\
  \bibinfo {author} {\bibfnamefont {J.~A.}\ \bibnamefont {Pons}},\ }\href
  {\doibase 10.1103/PhysRevC.59.2888} {\bibfield  {journal} {\bibinfo
  {journal} {Phys. Rev.}\ }\textbf {\bibinfo {volume} {C59}},\ \bibinfo {pages}
  {2888} (\bibinfo {year} {1999})}\BibitemShut {NoStop}%
\bibitem [{\citenamefont {Navarro}\ \emph {et~al.}(1999)\citenamefont
  {Navarro}, \citenamefont {Hernandez},\ and\ \citenamefont
  {Vautherin}}]{Navarro:1999jn}%
  \BibitemOpen
  \bibfield  {author} {\bibinfo {author} {\bibfnamefont {J.}~\bibnamefont
  {Navarro}}, \bibinfo {author} {\bibfnamefont {E.~S.}\ \bibnamefont
  {Hernandez}}, \ and\ \bibinfo {author} {\bibfnamefont {D.}~\bibnamefont
  {Vautherin}},\ }\href {\doibase 10.1103/PhysRevC.60.045801} {\bibfield
  {journal} {\bibinfo  {journal} {Phys. Rev.}\ }\textbf {\bibinfo {volume}
  {C60}},\ \bibinfo {pages} {045801} (\bibinfo {year} {1999})}\BibitemShut
  {NoStop}%
\bibitem [{\citenamefont {Margueron}\ \emph {et~al.}(2004)\citenamefont
  {Margueron}, \citenamefont {Navarro},\ and\ \citenamefont
  {Blottiau}}]{Margueron:2004sr}%
  \BibitemOpen
  \bibfield  {author} {\bibinfo {author} {\bibfnamefont {J.}~\bibnamefont
  {Margueron}}, \bibinfo {author} {\bibfnamefont {J.}~\bibnamefont {Navarro}},
  \ and\ \bibinfo {author} {\bibfnamefont {P.}~\bibnamefont {Blottiau}},\
  }\href {\doibase 10.1103/PhysRevC.70.028801} {\bibfield  {journal} {\bibinfo
  {journal} {Phys. Rev.}\ }\textbf {\bibinfo {volume} {C70}},\ \bibinfo {pages}
  {028801} (\bibinfo {year} {2004})}\BibitemShut {NoStop}%
\bibitem [{\citenamefont {Horowitz}\ and\ \citenamefont
  {Schwenk}(2006)}]{Horowitz:2006pj}%
  \BibitemOpen
  \bibfield  {author} {\bibinfo {author} {\bibfnamefont {C.~J.}\ \bibnamefont
  {Horowitz}}\ and\ \bibinfo {author} {\bibfnamefont {A.}~\bibnamefont
  {Schwenk}},\ }\href {\doibase 10.1016/j.physletb.2006.09.042} {\bibfield
  {journal} {\bibinfo  {journal} {Phys. Lett.}\ }\textbf {\bibinfo {volume}
  {B642}},\ \bibinfo {pages} {326} (\bibinfo {year} {2006})}\BibitemShut
  {NoStop}%
\bibitem [{\citenamefont {Horowitz}\ \emph {et~al.}(2017)\citenamefont
  {Horowitz}, \citenamefont {Caballero}, \citenamefont {Lin}, \citenamefont
  {O'Connor},\ and\ \citenamefont {Schwenk}}]{Horowitz:2016gul}%
  \BibitemOpen
  \bibfield  {author} {\bibinfo {author} {\bibfnamefont {C.~J.}\ \bibnamefont
  {Horowitz}}, \bibinfo {author} {\bibfnamefont {O.~L.}\ \bibnamefont
  {Caballero}}, \bibinfo {author} {\bibfnamefont {Z.}~\bibnamefont {Lin}},
  \bibinfo {author} {\bibfnamefont {E.}~\bibnamefont {O'Connor}}, \ and\
  \bibinfo {author} {\bibfnamefont {A.}~\bibnamefont {Schwenk}},\ }\href
  {\doibase 10.1103/PhysRevC.95.025801} {\bibfield  {journal} {\bibinfo
  {journal} {Phys. Rev.}\ }\textbf {\bibinfo {volume} {C95}},\ \bibinfo {pages}
  {025801} (\bibinfo {year} {2017})}\BibitemShut {NoStop}%
\bibitem [{\citenamefont {Leinson}\ and\ \citenamefont
  {Perez}(2001)}]{Leinson:2001ei}%
  \BibitemOpen
  \bibfield  {author} {\bibinfo {author} {\bibfnamefont {L.~B.}\ \bibnamefont
  {Leinson}}\ and\ \bibinfo {author} {\bibfnamefont {A.}~\bibnamefont
  {Perez}},\ }\href {\doibase 10.1016/S0370-2693(01)01042-5,
  10.1016/S0370-2693(01)01301-6} {\bibfield  {journal} {\bibinfo  {journal}
  {Phys. Lett.}\ }\textbf {\bibinfo {volume} {B518}},\ \bibinfo {pages} {15}
  (\bibinfo {year} {2001})},\ \bibinfo {note} {[Erratum: Phys.
  Lett.B522,358(2001)]}\BibitemShut {NoStop}%
\bibitem [{\citenamefont {Leinson}(2002)}]{Leinson:2002bw}%
  \BibitemOpen
  \bibfield  {author} {\bibinfo {author} {\bibfnamefont {L.~B.}\ \bibnamefont
  {Leinson}},\ }\href {\doibase 10.1016/S0375-9474(02)00991-0} {\bibfield
  {journal} {\bibinfo  {journal} {Nucl. Phys.}\ }\textbf {\bibinfo {volume}
  {A707}},\ \bibinfo {pages} {543} (\bibinfo {year} {2002})}\BibitemShut
  {NoStop}%
\bibitem [{\citenamefont {Roberts}\ and\ \citenamefont
  {Reddy}(2017)}]{Roberts:2016mwj}%
  \BibitemOpen
  \bibfield  {author} {\bibinfo {author} {\bibfnamefont {L.~F.}\ \bibnamefont
  {Roberts}}\ and\ \bibinfo {author} {\bibfnamefont {S.}~\bibnamefont
  {Reddy}},\ }\href {\doibase 10.1103/PhysRevC.95.045807} {\bibfield  {journal}
  {\bibinfo  {journal} {Phys. Rev.}\ }\textbf {\bibinfo {volume} {C95}},\
  \bibinfo {pages} {045807} (\bibinfo {year} {2017})}\BibitemShut {NoStop}%
\bibitem [{\citenamefont {O'Connor}(2015)}]{OConnor:2014sgn}%
  \BibitemOpen
  \bibfield  {author} {\bibinfo {author} {\bibfnamefont {E.}~\bibnamefont
  {O'Connor}},\ }\href {\doibase 10.1088/0067-0049/219/2/24} {\bibfield
  {journal} {\bibinfo  {journal} {Astrophys. J. Suppl.}\ }\textbf {\bibinfo
  {volume} {219}},\ \bibinfo {pages} {24} (\bibinfo {year} {2015})}\BibitemShut
  {NoStop}%
\bibitem [{\citenamefont {Pons}\ \emph {et~al.}(1999)\citenamefont {Pons},
  \citenamefont {Reddy}, \citenamefont {Prakash}, \citenamefont {Lattimer},\
  and\ \citenamefont {Miralles}}]{Pons:1998mm}%
  \BibitemOpen
  \bibfield  {author} {\bibinfo {author} {\bibfnamefont {J.}~\bibnamefont
  {Pons}}, \bibinfo {author} {\bibfnamefont {S.}~\bibnamefont {Reddy}},
  \bibinfo {author} {\bibfnamefont {M.}~\bibnamefont {Prakash}}, \bibinfo
  {author} {\bibfnamefont {J.}~\bibnamefont {Lattimer}}, \ and\ \bibinfo
  {author} {\bibfnamefont {J.}~\bibnamefont {Miralles}},\ }\href {\doibase
  10.1086/306889} {\bibfield  {journal} {\bibinfo  {journal} {Astrophys. J.}\
  }\textbf {\bibinfo {volume} {513}},\ \bibinfo {pages} {780} (\bibinfo {year}
  {1999})},\ \Eprint {http://arxiv.org/abs/astro-ph/9807040}
  {arXiv:astro-ph/9807040} \BibitemShut {NoStop}%
\bibitem [{\citenamefont {Buras}\ \emph {et~al.}(2006)\citenamefont {Buras},
  \citenamefont {Janka}, \citenamefont {Rampp},\ and\ \citenamefont
  {Kifonidis}}]{Buras:2005tb}%
  \BibitemOpen
  \bibfield  {author} {\bibinfo {author} {\bibfnamefont {R.}~\bibnamefont
  {Buras}}, \bibinfo {author} {\bibfnamefont {H.-T.}\ \bibnamefont {Janka}},
  \bibinfo {author} {\bibfnamefont {M.}~\bibnamefont {Rampp}}, \ and\ \bibinfo
  {author} {\bibfnamefont {K.}~\bibnamefont {Kifonidis}},\ }\href {\doibase
  10.1051/0004-6361:20054654} {\bibfield  {journal} {\bibinfo  {journal}
  {Astron. Astrophys.}\ }\textbf {\bibinfo {volume} {457}},\ \bibinfo {pages}
  {281} (\bibinfo {year} {2006})},\ \Eprint
  {http://arxiv.org/abs/astro-ph/0512189} {arXiv:astro-ph/0512189} \BibitemShut
  {NoStop}%
\bibitem [{\citenamefont {H{\"u}depohl}\ \emph {et~al.}(2010)\citenamefont
  {H{\"u}depohl}, \citenamefont {M{\"u}ller}, \citenamefont {Janka},
  \citenamefont {Marek},\ and\ \citenamefont {Raffelt}}]{Huedepohl:2009wh}%
  \BibitemOpen
  \bibfield  {author} {\bibinfo {author} {\bibfnamefont {L.}~\bibnamefont
  {H{\"u}depohl}}, \bibinfo {author} {\bibfnamefont {B.}~\bibnamefont
  {M{\"u}ller}}, \bibinfo {author} {\bibfnamefont {H.-T.}\ \bibnamefont
  {Janka}}, \bibinfo {author} {\bibfnamefont {A.}~\bibnamefont {Marek}}, \ and\
  \bibinfo {author} {\bibfnamefont {G.}~\bibnamefont {Raffelt}},\ }\href
  {\doibase 10.1103/PhysRevLett.104.251101} {\bibfield  {journal} {\bibinfo
  {journal} {Phys. Rev. Lett.}\ }\textbf {\bibinfo {volume} {104}},\ \bibinfo
  {pages} {251101} (\bibinfo {year} {2010})},\ \bibinfo {note} {[Erratum:
  Phys.Rev.Lett. 105, 249901 (2010)]}\BibitemShut {NoStop}%
\bibitem [{\citenamefont {{Fuller}}\ \emph {et~al.}(1982)\citenamefont
  {{Fuller}}, \citenamefont {{Fowler}},\ and\ \citenamefont
  {{Newman}}}]{FFN_1982b}%
  \BibitemOpen
  \bibfield  {author} {\bibinfo {author} {\bibfnamefont {G.~M.}\ \bibnamefont
  {{Fuller}}}, \bibinfo {author} {\bibfnamefont {W.~A.}\ \bibnamefont
  {{Fowler}}}, \ and\ \bibinfo {author} {\bibfnamefont {M.~J.}\ \bibnamefont
  {{Newman}}},\ }\href {\doibase 10.1086/190779} {\bibfield  {journal}
  {\bibinfo  {journal} {Astrophys. J. Suppl.}\ }\textbf {\bibinfo {volume}
  {48}},\ \bibinfo {pages} {279} (\bibinfo {year} {1982})}\BibitemShut
  {NoStop}%
\bibitem [{\citenamefont {Langanke}\ and\ \citenamefont
  {Martinez-Pinedo}(2001)}]{LMP_ADNDT_2001}%
  \BibitemOpen
  \bibfield  {author} {\bibinfo {author} {\bibfnamefont {K.}~\bibnamefont
  {Langanke}}\ and\ \bibinfo {author} {\bibfnamefont {G.}~\bibnamefont
  {Martinez-Pinedo}},\ }\href {\doibase 10.1006/adnd.2001.0865} {\bibfield
  {journal} {\bibinfo  {journal} {Atom. Data Nucl. Data Tabl.}\ }\textbf
  {\bibinfo {volume} {79}},\ \bibinfo {pages} {1} (\bibinfo {year}
  {2001})}\BibitemShut {NoStop}%
\bibitem [{\citenamefont {Langanke}\ and\ \citenamefont
  {Martinez-Pinedo}(2003)}]{Langanke2002}%
  \BibitemOpen
  \bibfield  {author} {\bibinfo {author} {\bibfnamefont {K.}~\bibnamefont
  {Langanke}}\ and\ \bibinfo {author} {\bibfnamefont {G.}~\bibnamefont
  {Martinez-Pinedo}},\ }\href {\doibase 10.1103/RevModPhys.75.819} {\bibfield
  {journal} {\bibinfo  {journal} {Rev. Mod. Phys.}\ }\textbf {\bibinfo {volume}
  {75}},\ \bibinfo {pages} {819} (\bibinfo {year} {2003})}\BibitemShut
  {NoStop}%
\bibitem [{\citenamefont {Oda}\ \emph {et~al.}(1994)\citenamefont {Oda},
  \citenamefont {Hino}, \citenamefont {Muto}, \citenamefont {Takahara},\ and\
  \citenamefont {Sato}}]{Oda1994}%
  \BibitemOpen
  \bibfield  {author} {\bibinfo {author} {\bibfnamefont {T.}~\bibnamefont
  {Oda}}, \bibinfo {author} {\bibfnamefont {M.}~\bibnamefont {Hino}}, \bibinfo
  {author} {\bibfnamefont {K.}~\bibnamefont {Muto}}, \bibinfo {author}
  {\bibfnamefont {M.}~\bibnamefont {Takahara}}, \ and\ \bibinfo {author}
  {\bibfnamefont {K.}~\bibnamefont {Sato}},\ }\href {\doibase
  10.1006/adnd.1994.1007} {\bibfield  {journal} {\bibinfo  {journal} {Atom.
  Data Nucl. Data Tabl.}\ }\textbf {\bibinfo {volume} {56}},\ \bibinfo {pages}
  {231} (\bibinfo {year} {1994})}\BibitemShut {NoStop}%
\bibitem [{\citenamefont {Pruet}\ and\ \citenamefont
  {Fuller}(2003)}]{Pruet2003}%
  \BibitemOpen
  \bibfield  {author} {\bibinfo {author} {\bibfnamefont {J.}~\bibnamefont
  {Pruet}}\ and\ \bibinfo {author} {\bibfnamefont {G.~M.}\ \bibnamefont
  {Fuller}},\ }\href {\doibase 10.1086/376753} {\bibfield  {journal} {\bibinfo
  {journal} {Astrophys. J. Suppl.}\ }\textbf {\bibinfo {volume} {149}},\
  \bibinfo {pages} {189} (\bibinfo {year} {2003})}\BibitemShut {NoStop}%
\bibitem [{\citenamefont {Juodagalvis}\ \emph {et~al.}(2010)\citenamefont
  {Juodagalvis}, \citenamefont {Langanke}, \citenamefont {Hix}, \citenamefont
  {Martínez-Pinedo},\ and\ \citenamefont {Sampaio}}]{Juodagalvis_NPA_2010}%
  \BibitemOpen
  \bibfield  {author} {\bibinfo {author} {\bibfnamefont {A.}~\bibnamefont
  {Juodagalvis}}, \bibinfo {author} {\bibfnamefont {K.}~\bibnamefont
  {Langanke}}, \bibinfo {author} {\bibfnamefont {W.}~\bibnamefont {Hix}},
  \bibinfo {author} {\bibfnamefont {G.}~\bibnamefont {Martínez-Pinedo}}, \
  and\ \bibinfo {author} {\bibfnamefont {J.}~\bibnamefont {Sampaio}},\ }\href
  {\doibase https://doi.org/10.1016/j.nuclphysa.2010.09.012} {\bibfield
  {journal} {\bibinfo  {journal} {Nuclear Physics A}\ }\textbf {\bibinfo
  {volume} {848}},\ \bibinfo {pages} {454 } (\bibinfo {year}
  {2010})}\BibitemShut {NoStop}%
\bibitem [{\citenamefont {Typel}\ \emph {et~al.}(2015)\citenamefont {Typel},
  \citenamefont {Oertel},\ and\ \citenamefont {Klähn}}]{Typel:2013rza}%
  \BibitemOpen
  \bibfield  {author} {\bibinfo {author} {\bibfnamefont {S.}~\bibnamefont
  {Typel}}, \bibinfo {author} {\bibfnamefont {M.}~\bibnamefont {Oertel}}, \
  and\ \bibinfo {author} {\bibfnamefont {T.}~\bibnamefont {Klähn}},\ }\href
  {\doibase 10.1134/S1063779615040061} {\bibfield  {journal} {\bibinfo
  {journal} {Phys. Part. Nucl.}\ }\textbf {\bibinfo {volume} {46}},\ \bibinfo
  {pages} {633} (\bibinfo {year} {2015})}\BibitemShut {NoStop}%
\bibitem [{\citenamefont {Hernandez}\ \emph {et~al.}(1999)\citenamefont
  {Hernandez}, \citenamefont {Navarro},\ and\ \citenamefont
  {Polls}}]{Hernandez:1999zz}%
  \BibitemOpen
  \bibfield  {author} {\bibinfo {author} {\bibfnamefont {E.~S.}\ \bibnamefont
  {Hernandez}}, \bibinfo {author} {\bibfnamefont {J.}~\bibnamefont {Navarro}},
  \ and\ \bibinfo {author} {\bibfnamefont {A.}~\bibnamefont {Polls}},\ }\href
  {\doibase 10.1016/S0375-9474(99)00363-2} {\bibfield  {journal} {\bibinfo
  {journal} {Nucl. Phys.}\ }\textbf {\bibinfo {volume} {A658}},\ \bibinfo
  {pages} {327} (\bibinfo {year} {1999})}\BibitemShut {NoStop}%
\bibitem [{\citenamefont {Dzhioev}\ and\ \citenamefont
  {Martínez-Pinedo}(2018)}]{Dzhioev:2018ovi}%
  \BibitemOpen
  \bibfield  {author} {\bibinfo {author} {\bibfnamefont {A.~A.}\ \bibnamefont
  {Dzhioev}}\ and\ \bibinfo {author} {\bibfnamefont {G.}~\bibnamefont
  {Martínez-Pinedo}},\ }\bibfield  {booktitle} {\emph {\bibinfo {booktitle}
  {{Proceedings, International Conference Nuclear Structure and Related Topics
  (NRST18): Burgas, Bulgaria, June 3-9, 2018}}},\ }\href {\doibase
  10.1051/epjconf/201819402006} {\bibfield  {journal} {\bibinfo  {journal} {EPJ
  Web Conf.}\ }\textbf {\bibinfo {volume} {194}},\ \bibinfo {pages} {02006}
  (\bibinfo {year} {2018})}\BibitemShut {NoStop}%
\bibitem [{\citenamefont {Sedrakian}\ and\ \citenamefont
  {Dieperink}(1999)}]{Sedrakian:1999jh}%
  \BibitemOpen
  \bibfield  {author} {\bibinfo {author} {\bibfnamefont {A.}~\bibnamefont
  {Sedrakian}}\ and\ \bibinfo {author} {\bibfnamefont {A.}~\bibnamefont
  {Dieperink}},\ }\href {\doibase 10.1016/S0370-2693(99)00989-2} {\bibfield
  {journal} {\bibinfo  {journal} {Phys. Lett.}\ }\textbf {\bibinfo {volume}
  {B463}},\ \bibinfo {pages} {145} (\bibinfo {year} {1999})}\BibitemShut
  {NoStop}%
\bibitem [{\citenamefont {Schmitt}\ \emph {et~al.}(2006)\citenamefont
  {Schmitt}, \citenamefont {Shovkovy},\ and\ \citenamefont
  {Wang}}]{Schmitt:2005wg}%
  \BibitemOpen
  \bibfield  {author} {\bibinfo {author} {\bibfnamefont {A.}~\bibnamefont
  {Schmitt}}, \bibinfo {author} {\bibfnamefont {I.~A.}\ \bibnamefont
  {Shovkovy}}, \ and\ \bibinfo {author} {\bibfnamefont {Q.}~\bibnamefont
  {Wang}},\ }\href {\doibase 10.1103/PhysRevD.73.034012} {\bibfield  {journal}
  {\bibinfo  {journal} {Phys. Rev.}\ }\textbf {\bibinfo {volume} {D73}},\
  \bibinfo {pages} {034012} (\bibinfo {year} {2006})}\BibitemShut {NoStop}%
\bibitem [{\citenamefont {Hempel}(2015)}]{Hempel:2014ssa}%
  \BibitemOpen
  \bibfield  {author} {\bibinfo {author} {\bibfnamefont {M.}~\bibnamefont
  {Hempel}},\ }\href {\doibase 10.1103/PhysRevC.91.055807} {\bibfield
  {journal} {\bibinfo  {journal} {Phys. Rev.}\ }\textbf {\bibinfo {volume}
  {C91}},\ \bibinfo {pages} {055807} (\bibinfo {year} {2015})}\BibitemShut
  {NoStop}%
\bibitem [{\citenamefont {Ducoin}\ \emph {et~al.}(2006)\citenamefont {Ducoin},
  \citenamefont {Chomaz},\ and\ \citenamefont {Gulminelli}}]{Ducoin:2005aa}%
  \BibitemOpen
  \bibfield  {author} {\bibinfo {author} {\bibfnamefont {C.}~\bibnamefont
  {Ducoin}}, \bibinfo {author} {\bibfnamefont {P.}~\bibnamefont {Chomaz}}, \
  and\ \bibinfo {author} {\bibfnamefont {F.}~\bibnamefont {Gulminelli}},\
  }\href {\doibase 10.1016/j.nuclphysa.2006.03.005} {\bibfield  {journal}
  {\bibinfo  {journal} {Nucl. Phys.}\ }\textbf {\bibinfo {volume} {A771}},\
  \bibinfo {pages} {68} (\bibinfo {year} {2006})},\ \Eprint
  {http://arxiv.org/abs/nucl-th/0512029} {arXiv:nucl-th/0512029 [nucl-th]}
  \BibitemShut {NoStop}%
\bibitem [{\citenamefont {Horowitz}\ \emph {et~al.}(2012)\citenamefont
  {Horowitz}, \citenamefont {Shen}, \citenamefont {O'Connor},\ and\
  \citenamefont {Ott}}]{Horowitz:2012us}%
  \BibitemOpen
  \bibfield  {author} {\bibinfo {author} {\bibfnamefont {C.~J.}\ \bibnamefont
  {Horowitz}}, \bibinfo {author} {\bibfnamefont {G.}~\bibnamefont {Shen}},
  \bibinfo {author} {\bibfnamefont {E.}~\bibnamefont {O'Connor}}, \ and\
  \bibinfo {author} {\bibfnamefont {C.~D.}\ \bibnamefont {Ott}},\ }\href
  {\doibase 10.1103/PhysRevC.86.065806} {\bibfield  {journal} {\bibinfo
  {journal} {Phys. Rev.}\ }\textbf {\bibinfo {volume} {C86}},\ \bibinfo {pages}
  {065806} (\bibinfo {year} {2012})}\BibitemShut {NoStop}%
\bibitem [{\citenamefont {Margueron}(2001)}]{Margueronphd}%
  \BibitemOpen
  \bibfield  {author} {\bibinfo {author} {\bibfnamefont {J.}~\bibnamefont
  {Margueron}},\ }\emph {\bibinfo {title} {{Effects du milieu sur la
  propagation des neutrinos dans la matière nucléaire}}},\ \href@noop {}
  {Ph.D. thesis},\ \bibinfo  {school} {Paris 11} (\bibinfo {year}
  {2001})\BibitemShut {NoStop}%
\bibitem [{\citenamefont {Pastore}\ \emph {et~al.}(2014)\citenamefont
  {Pastore}, \citenamefont {Davesne},\ and\ \citenamefont
  {Navarro}}]{Pastore:2014aia}%
  \BibitemOpen
  \bibfield  {author} {\bibinfo {author} {\bibfnamefont {A.}~\bibnamefont
  {Pastore}}, \bibinfo {author} {\bibfnamefont {D.}~\bibnamefont {Davesne}}, \
  and\ \bibinfo {author} {\bibfnamefont {J.}~\bibnamefont {Navarro}},\ }\href
  {\doibase 10.1016/j.physrep.2014.11.002} {\bibfield  {journal} {\bibinfo
  {journal} {Phys. Rept.}\ }\textbf {\bibinfo {volume} {563}},\ \bibinfo
  {pages} {1} (\bibinfo {year} {2014})},\ \Eprint
  {http://arxiv.org/abs/1412.2339} {arXiv:1412.2339 [nucl-th]} \BibitemShut
  {NoStop}%
\bibitem [{\citenamefont {Rapp}\ \emph {et~al.}(1998)\citenamefont {Rapp},
  \citenamefont {Urban}, \citenamefont {Buballa},\ and\ \citenamefont
  {Wambach}}]{Rapp:1997ei}%
  \BibitemOpen
  \bibfield  {author} {\bibinfo {author} {\bibfnamefont {R.}~\bibnamefont
  {Rapp}}, \bibinfo {author} {\bibfnamefont {M.}~\bibnamefont {Urban}},
  \bibinfo {author} {\bibfnamefont {M.}~\bibnamefont {Buballa}}, \ and\
  \bibinfo {author} {\bibfnamefont {J.}~\bibnamefont {Wambach}},\ }\href
  {\doibase 10.1016/S0370-2693(97)01360-9} {\bibfield  {journal} {\bibinfo
  {journal} {Phys. Lett.}\ }\textbf {\bibinfo {volume} {B417}},\ \bibinfo
  {pages} {1} (\bibinfo {year} {1998})}\BibitemShut {NoStop}%
\bibitem [{\citenamefont {Margueron}\ and\ \citenamefont
  {Sagawa}(2009)}]{Margueron:2009rn}%
  \BibitemOpen
  \bibfield  {author} {\bibinfo {author} {\bibfnamefont {J.}~\bibnamefont
  {Margueron}}\ and\ \bibinfo {author} {\bibfnamefont {H.}~\bibnamefont
  {Sagawa}},\ }\href {\doibase 10.1088/0954-3899/36/12/125102} {\bibfield
  {journal} {\bibinfo  {journal} {J. Phys.}\ }\textbf {\bibinfo {volume}
  {G36}},\ \bibinfo {pages} {125102} (\bibinfo {year} {2009})},\ \Eprint
  {http://arxiv.org/abs/0905.1931} {arXiv:0905.1931 [nucl-th]} \BibitemShut
  {NoStop}%
\bibitem [{\citenamefont {Davesne}\ \emph {et~al.}(2019)\citenamefont
  {Davesne}, \citenamefont {Pastore},\ and\ \citenamefont
  {Navarro}}]{Davesne:2019ytl}%
  \BibitemOpen
  \bibfield  {author} {\bibinfo {author} {\bibfnamefont {D.}~\bibnamefont
  {Davesne}}, \bibinfo {author} {\bibfnamefont {A.}~\bibnamefont {Pastore}}, \
  and\ \bibinfo {author} {\bibfnamefont {J.}~\bibnamefont {Navarro}},\
  }\href@noop {} {\  (\bibinfo {year} {2019})},\ \Eprint
  {http://arxiv.org/abs/1905.12049} {arXiv:1905.12049 [nucl-th]} \BibitemShut
  {NoStop}%
\bibitem [{\citenamefont {Gulminelli}\ and\ \citenamefont
  {Raduta}(2015)}]{Gulminelli_PRC_2015}%
  \BibitemOpen
  \bibfield  {author} {\bibinfo {author} {\bibfnamefont {F.}~\bibnamefont
  {Gulminelli}}\ and\ \bibinfo {author} {\bibfnamefont {A.~R.}\ \bibnamefont
  {Raduta}},\ }\href {\doibase 10.1103/PhysRevC.92.055803} {\bibfield
  {journal} {\bibinfo  {journal} {Phys. Rev. C}\ }\textbf {\bibinfo {volume}
  {92}},\ \bibinfo {pages} {055803} (\bibinfo {year} {2015})}\BibitemShut
  {NoStop}%
\bibitem [{\citenamefont {Raduta}\ and\ \citenamefont
  {Gulminelli}(2019)}]{Raduta_2019}%
  \BibitemOpen
  \bibfield  {author} {\bibinfo {author} {\bibfnamefont {A.~R.}\ \bibnamefont
  {Raduta}}\ and\ \bibinfo {author} {\bibfnamefont {F.}~\bibnamefont
  {Gulminelli}},\ }\href {\doibase 10.1016/j.nuclphysa.2018.11.003} {\bibfield
  {journal} {\bibinfo  {journal} {Nucl. Phys.}\ }\textbf {\bibinfo {volume}
  {A983}},\ \bibinfo {pages} {252} (\bibinfo {year} {2019})}\BibitemShut
  {NoStop}%
\bibitem [{\citenamefont {Chabanat}\ \emph {et~al.}(1998)\citenamefont
  {Chabanat}, \citenamefont {Bonche}, \citenamefont {Haensel}, \citenamefont
  {Meyer},\ and\ \citenamefont {Schaeffer}}]{SLy4}%
  \BibitemOpen
  \bibfield  {author} {\bibinfo {author} {\bibfnamefont {E.}~\bibnamefont
  {Chabanat}}, \bibinfo {author} {\bibfnamefont {P.}~\bibnamefont {Bonche}},
  \bibinfo {author} {\bibfnamefont {P.}~\bibnamefont {Haensel}}, \bibinfo
  {author} {\bibfnamefont {J.}~\bibnamefont {Meyer}}, \ and\ \bibinfo {author}
  {\bibfnamefont {R.}~\bibnamefont {Schaeffer}},\ }\href {\doibase
  https://doi.org/10.1016/S0375-9474(98)00180-8} {\bibfield  {journal}
  {\bibinfo  {journal} {Nuclear Physics A}\ }\textbf {\bibinfo {volume}
  {635}},\ \bibinfo {pages} {231 } (\bibinfo {year} {1998})}\BibitemShut
  {NoStop}%
\bibitem [{\citenamefont {Hempel}\ and\ \citenamefont
  {Schaffner-Bielich}(2010)}]{Hempel_NPA_2010}%
  \BibitemOpen
  \bibfield  {author} {\bibinfo {author} {\bibfnamefont {M.}~\bibnamefont
  {Hempel}}\ and\ \bibinfo {author} {\bibfnamefont {J.}~\bibnamefont
  {Schaffner-Bielich}},\ }\href {\doibase
  https://doi.org/10.1016/j.nuclphysa.2010.02.010} {\bibfield  {journal}
  {\bibinfo  {journal} {Nuclear Physics A}\ }\textbf {\bibinfo {volume}
  {837}},\ \bibinfo {pages} {210 } (\bibinfo {year} {2010})}\BibitemShut
  {NoStop}%
\bibitem [{\citenamefont {Typel}\ \emph {et~al.}(2010)\citenamefont {Typel},
  \citenamefont {R\"opke}, \citenamefont {Kl\"ahn}, \citenamefont {Blaschke},\
  and\ \citenamefont {Wolter}}]{DD2}%
  \BibitemOpen
  \bibfield  {author} {\bibinfo {author} {\bibfnamefont {S.}~\bibnamefont
  {Typel}}, \bibinfo {author} {\bibfnamefont {G.}~\bibnamefont {R\"opke}},
  \bibinfo {author} {\bibfnamefont {T.}~\bibnamefont {Kl\"ahn}}, \bibinfo
  {author} {\bibfnamefont {D.}~\bibnamefont {Blaschke}}, \ and\ \bibinfo
  {author} {\bibfnamefont {H.~H.}\ \bibnamefont {Wolter}},\ }\href {\doibase
  10.1103/PhysRevC.81.015803} {\bibfield  {journal} {\bibinfo  {journal} {Phys.
  Rev. C}\ }\textbf {\bibinfo {volume} {81}},\ \bibinfo {pages} {015803}
  (\bibinfo {year} {2010})}\BibitemShut {NoStop}%
\bibitem [{\citenamefont {Hempel}\ \emph {et~al.}(2012)\citenamefont {Hempel},
  \citenamefont {Fischer}, \citenamefont {Schaffner-Bielich},\ and\
  \citenamefont {Liebendörfer}}]{Hempel_ApJ_2012}%
  \BibitemOpen
  \bibfield  {author} {\bibinfo {author} {\bibfnamefont {M.}~\bibnamefont
  {Hempel}}, \bibinfo {author} {\bibfnamefont {T.}~\bibnamefont {Fischer}},
  \bibinfo {author} {\bibfnamefont {J.}~\bibnamefont {Schaffner-Bielich}}, \
  and\ \bibinfo {author} {\bibfnamefont {M.}~\bibnamefont {Liebendörfer}},\
  }\href {http://stacks.iop.org/0004-637X/748/i=1/a=70} {\bibfield  {journal}
  {\bibinfo  {journal} {The Astrophysical Journal}\ }\textbf {\bibinfo {volume}
  {748}},\ \bibinfo {pages} {70} (\bibinfo {year} {2012})}\BibitemShut
  {NoStop}%
\bibitem [{\citenamefont {Demorest}\ \emph {et~al.}(2010)\citenamefont
  {Demorest}, \citenamefont {Pennucci}, \citenamefont {Ransom}, \citenamefont
  {Roberts},\ and\ \citenamefont {Hessels}}]{Demorest_10}%
  \BibitemOpen
  \bibfield  {author} {\bibinfo {author} {\bibfnamefont {P.}~\bibnamefont
  {Demorest}}, \bibinfo {author} {\bibfnamefont {T.}~\bibnamefont {Pennucci}},
  \bibinfo {author} {\bibfnamefont {S.}~\bibnamefont {Ransom}}, \bibinfo
  {author} {\bibfnamefont {M.}~\bibnamefont {Roberts}}, \ and\ \bibinfo
  {author} {\bibfnamefont {J.}~\bibnamefont {Hessels}},\ }\href {\doibase
  10.1038/nature09466} {\bibfield  {journal} {\bibinfo  {journal} {Nature}\
  }\textbf {\bibinfo {volume} {467}},\ \bibinfo {pages} {1081} (\bibinfo {year}
  {2010})},\ \Eprint {http://arxiv.org/abs/1010.5788} {arXiv:1010.5788
  [astro-ph.HE]} \BibitemShut {NoStop}%
\bibitem [{\citenamefont {{Antoniadis}}\ \emph {et~al.}(2013)\citenamefont
  {{Antoniadis}}, \citenamefont {{Freire}}, \citenamefont {{Wex}},
  \citenamefont {{Tauris}}, \citenamefont {{Lynch}}, \citenamefont {{van
  Kerkwijk}}, \citenamefont {{Kramer}}, \citenamefont {{Bassa}}, \citenamefont
  {{Dhillon}}, \citenamefont {{Driebe}}, \citenamefont {{Hessels}},
  \citenamefont {{Kaspi}}, \citenamefont {{Kondratiev}}, \citenamefont
  {{Langer}}, \citenamefont {{Marsh}}, \citenamefont {{McLaughlin}},
  \citenamefont {{Pennucci}}, \citenamefont {{Ransom}}, \citenamefont
  {{Stairs}}, \citenamefont {{van Leeuwen}}, \citenamefont {{Verbiest}},\ and\
  \citenamefont {{Whelan}}}]{Antoniadis_13}%
  \BibitemOpen
  \bibfield  {author} {\bibinfo {author} {\bibfnamefont {J.}~\bibnamefont
  {{Antoniadis}}}, \bibinfo {author} {\bibfnamefont {P.~C.~C.}\ \bibnamefont
  {{Freire}}}, \bibinfo {author} {\bibfnamefont {N.}~\bibnamefont {{Wex}}},
  \bibinfo {author} {\bibfnamefont {T.~M.}\ \bibnamefont {{Tauris}}}, \bibinfo
  {author} {\bibfnamefont {R.~S.}\ \bibnamefont {{Lynch}}}, \bibinfo {author}
  {\bibfnamefont {M.~H.}\ \bibnamefont {{van Kerkwijk}}}, \bibinfo {author}
  {\bibfnamefont {M.}~\bibnamefont {{Kramer}}}, \bibinfo {author}
  {\bibfnamefont {C.}~\bibnamefont {{Bassa}}}, \bibinfo {author} {\bibfnamefont
  {V.~S.}\ \bibnamefont {{Dhillon}}}, \bibinfo {author} {\bibfnamefont
  {T.}~\bibnamefont {{Driebe}}}, \bibinfo {author} {\bibfnamefont {J.~W.~T.}\
  \bibnamefont {{Hessels}}}, \bibinfo {author} {\bibfnamefont {V.~M.}\
  \bibnamefont {{Kaspi}}}, \bibinfo {author} {\bibfnamefont {V.~I.}\
  \bibnamefont {{Kondratiev}}}, \bibinfo {author} {\bibfnamefont
  {N.}~\bibnamefont {{Langer}}}, \bibinfo {author} {\bibfnamefont {T.~R.}\
  \bibnamefont {{Marsh}}}, \bibinfo {author} {\bibfnamefont {M.~A.}\
  \bibnamefont {{McLaughlin}}}, \bibinfo {author} {\bibfnamefont {T.~T.}\
  \bibnamefont {{Pennucci}}}, \bibinfo {author} {\bibfnamefont {S.~M.}\
  \bibnamefont {{Ransom}}}, \bibinfo {author} {\bibfnamefont {I.~H.}\
  \bibnamefont {{Stairs}}}, \bibinfo {author} {\bibfnamefont {J.}~\bibnamefont
  {{van Leeuwen}}}, \bibinfo {author} {\bibfnamefont {J.~P.~W.}\ \bibnamefont
  {{Verbiest}}}, \ and\ \bibinfo {author} {\bibfnamefont {D.~G.}\ \bibnamefont
  {{Whelan}}},\ }\href {\doibase 10.1126/science.1233232} {\bibfield  {journal}
  {\bibinfo  {journal} {Science}\ }\textbf {\bibinfo {volume} {340}},\ \bibinfo
  {pages} {448} (\bibinfo {year} {2013})},\ \Eprint
  {http://arxiv.org/abs/1304.6875} {arXiv:1304.6875 [astro-ph.HE]} \BibitemShut
  {NoStop}%
\bibitem [{\citenamefont {Arzoumanian}\ \emph {et~al.}(2018)\citenamefont
  {Arzoumanian} \emph {et~al.}}]{Arzoumanian:2017puf}%
  \BibitemOpen
  \bibfield  {author} {\bibinfo {author} {\bibfnamefont {Z.}~\bibnamefont
  {Arzoumanian}} \emph {et~al.} (\bibinfo {collaboration} {NANOGrav}),\ }\href
  {\doibase 10.3847/1538-4365/aab5b0} {\bibfield  {journal} {\bibinfo
  {journal} {Astrophys. J. Suppl.}\ }\textbf {\bibinfo {volume} {235}},\
  \bibinfo {pages} {37} (\bibinfo {year} {2018})},\ \Eprint
  {http://arxiv.org/abs/1801.01837} {arXiv:1801.01837 [astro-ph.HE]}
  \BibitemShut {NoStop}%
\bibitem [{\citenamefont {Oertel}\ \emph {et~al.}(2017)\citenamefont {Oertel},
  \citenamefont {Hempel}, \citenamefont {Klähn},\ and\ \citenamefont
  {Typel}}]{Oertel:2016bki}%
  \BibitemOpen
  \bibfield  {author} {\bibinfo {author} {\bibfnamefont {M.}~\bibnamefont
  {Oertel}}, \bibinfo {author} {\bibfnamefont {M.}~\bibnamefont {Hempel}},
  \bibinfo {author} {\bibfnamefont {T.}~\bibnamefont {Klähn}}, \ and\ \bibinfo
  {author} {\bibfnamefont {S.}~\bibnamefont {Typel}},\ }\href {\doibase
  10.1103/RevModPhys.89.015007} {\bibfield  {journal} {\bibinfo  {journal}
  {Rev. Mod. Phys.}\ }\textbf {\bibinfo {volume} {89}},\ \bibinfo {pages}
  {015007} (\bibinfo {year} {2017})},\ \Eprint
  {http://arxiv.org/abs/1610.03361} {arXiv:1610.03361 [astro-ph.HE]}
  \BibitemShut {NoStop}%
\bibitem [{\citenamefont {Abbott}\ \emph {et~al.}(2018)\citenamefont {Abbott}
  \emph {et~al.}}]{abbott_18}%
  \BibitemOpen
  \bibfield  {author} {\bibinfo {author} {\bibfnamefont {B.~P.}\ \bibnamefont
  {Abbott}} \emph {et~al.} (\bibinfo {collaboration} {Virgo, LIGO
  Scientific}),\ }\href@noop {} {\  (\bibinfo {year} {2018})},\ \Eprint
  {http://arxiv.org/abs/1805.11581} {arXiv:1805.11581 [gr-qc]} \BibitemShut
  {NoStop}%
\bibitem [{\citenamefont {Endrizzi}\ \emph {et~al.}(2019)\citenamefont
  {Endrizzi}, \citenamefont {Perego}, \citenamefont {Fabbri}, \citenamefont
  {Branca}, \citenamefont {Radice}, \citenamefont {Bernuzzi}, \citenamefont
  {Giacomazzo}, \citenamefont {Pederiva},\ and\ \citenamefont
  {Lovato}}]{Endrizzi:2019trv}%
  \BibitemOpen
  \bibfield  {author} {\bibinfo {author} {\bibfnamefont {A.}~\bibnamefont
  {Endrizzi}}, \bibinfo {author} {\bibfnamefont {A.}~\bibnamefont {Perego}},
  \bibinfo {author} {\bibfnamefont {F.~M.}\ \bibnamefont {Fabbri}}, \bibinfo
  {author} {\bibfnamefont {L.}~\bibnamefont {Branca}}, \bibinfo {author}
  {\bibfnamefont {D.}~\bibnamefont {Radice}}, \bibinfo {author} {\bibfnamefont
  {S.}~\bibnamefont {Bernuzzi}}, \bibinfo {author} {\bibfnamefont
  {B.}~\bibnamefont {Giacomazzo}}, \bibinfo {author} {\bibfnamefont
  {F.}~\bibnamefont {Pederiva}}, \ and\ \bibinfo {author} {\bibfnamefont
  {A.}~\bibnamefont {Lovato}},\ }\href@noop {} {\  (\bibinfo {year} {2019})},\
  \Eprint {http://arxiv.org/abs/1908.04952} {arXiv:1908.04952 [astro-ph.HE]}
  \BibitemShut {NoStop}%
\bibitem [{\citenamefont {{Dimmelmeier}}\ \emph {et~al.}(2012)\citenamefont
  {{Dimmelmeier}}, \citenamefont {{Novak}},\ and\ \citenamefont
  {{Cerd{\'a}-Dur{\'a}n}}}]{coconut}%
  \BibitemOpen
  \bibfield  {author} {\bibinfo {author} {\bibfnamefont {H.}~\bibnamefont
  {{Dimmelmeier}}}, \bibinfo {author} {\bibfnamefont {J.}~\bibnamefont
  {{Novak}}}, \ and\ \bibinfo {author} {\bibfnamefont {P.}~\bibnamefont
  {{Cerd{\'a}-Dur{\'a}n}}},\ }\href@noop {} {\enquote {\bibinfo {title}
  {{CoCoNuT: General relativistic hydrodynamics code with dynamical space- time
  evolution}},}\ }\bibinfo {howpublished} {Astrophysics Source Code Library}
  (\bibinfo {year} {2012}),\ \Eprint {http://arxiv.org/abs/1202.012}
  {ascl:1202.012} \BibitemShut {NoStop}%
\bibitem [{\citenamefont {Dimmelmeier}\ \emph {et~al.}(2005)\citenamefont
  {Dimmelmeier}, \citenamefont {Novak}, \citenamefont {Font}, \citenamefont
  {Ib{\'a}{\~n}ez},\ and\ \citenamefont {M{\"u}ller}}]{dimmelmeier-05}%
  \BibitemOpen
  \bibfield  {author} {\bibinfo {author} {\bibfnamefont {H.}~\bibnamefont
  {Dimmelmeier}}, \bibinfo {author} {\bibfnamefont {J.}~\bibnamefont {Novak}},
  \bibinfo {author} {\bibfnamefont {J.~A.}\ \bibnamefont {Font}}, \bibinfo
  {author} {\bibfnamefont {J.~M.}\ \bibnamefont {Ib{\'a}{\~n}ez}}, \ and\
  \bibinfo {author} {\bibfnamefont {E.}~\bibnamefont {M{\"u}ller}},\ }\href
  {\doibase 10.1103/PhysRevD.71.064023} {\bibfield  {journal} {\bibinfo
  {journal} {Phys. Rev. D}\ }\textbf {\bibinfo {volume} {71}},\ \bibinfo
  {pages} {1} (\bibinfo {year} {2005})}\BibitemShut {NoStop}%
\bibitem [{\citenamefont {{M{\"u}ller}}\ and\ \citenamefont
  {{Janka}}(2015)}]{mueller-15}%
  \BibitemOpen
  \bibfield  {author} {\bibinfo {author} {\bibfnamefont {B.}~\bibnamefont
  {{M{\"u}ller}}}\ and\ \bibinfo {author} {\bibfnamefont {H.~T.}\ \bibnamefont
  {{Janka}}},\ }\href {\doibase 10.1093/mnras/stv101} {\bibfield  {journal}
  {\bibinfo  {journal} {\mnras}\ }\textbf {\bibinfo {volume} {448}},\ \bibinfo
  {pages} {2141} (\bibinfo {year} {2015})},\ \Eprint
  {http://arxiv.org/abs/1409.4783} {arXiv:1409.4783 [astro-ph.SR]} \BibitemShut
  {NoStop}%
\bibitem [{\citenamefont {{Woosley}}\ \emph {et~al.}(2002)\citenamefont
  {{Woosley}}, \citenamefont {{Heger}},\ and\ \citenamefont
  {{Weaver}}}]{woosley-02}%
  \BibitemOpen
  \bibfield  {author} {\bibinfo {author} {\bibfnamefont {S.~E.}\ \bibnamefont
  {{Woosley}}}, \bibinfo {author} {\bibfnamefont {A.}~\bibnamefont {{Heger}}},
  \ and\ \bibinfo {author} {\bibfnamefont {T.~A.}\ \bibnamefont {{Weaver}}},\
  }\href {\doibase 10.1103/RevModPhys.74.1015} {\bibfield  {journal} {\bibinfo
  {journal} {Rev. Mod. Phys.}\ }\textbf {\bibinfo {volume} {74}},\ \bibinfo
  {pages} {1015} (\bibinfo {year} {2002})}\BibitemShut {NoStop}%
\bibitem [{\citenamefont {Hix}\ \emph {et~al.}(2003)\citenamefont {Hix},
  \citenamefont {Messer}, \citenamefont {Mezzacappa}, \citenamefont
  {Liebend{\"o}rfer}, \citenamefont {Sampaio}, \citenamefont {Langanke},
  \citenamefont {Dean},\ and\ \citenamefont {Martinez-Pinedo}}]{Hix2003}%
  \BibitemOpen
  \bibfield  {author} {\bibinfo {author} {\bibfnamefont {W.~R.}\ \bibnamefont
  {Hix}}, \bibinfo {author} {\bibfnamefont {O.~E.~B.}\ \bibnamefont {Messer}},
  \bibinfo {author} {\bibfnamefont {A.}~\bibnamefont {Mezzacappa}}, \bibinfo
  {author} {\bibfnamefont {M.}~\bibnamefont {Liebend{\"o}rfer}}, \bibinfo
  {author} {\bibfnamefont {J.}~\bibnamefont {Sampaio}}, \bibinfo {author}
  {\bibfnamefont {K.}~\bibnamefont {Langanke}}, \bibinfo {author}
  {\bibfnamefont {D.~J.}\ \bibnamefont {Dean}}, \ and\ \bibinfo {author}
  {\bibfnamefont {G.}~\bibnamefont {Martinez-Pinedo}},\ }\href {\doibase
  10.1103/PhysRevLett.91.201102} {\bibfield  {journal} {\bibinfo  {journal}
  {Phys. Rev. Lett.}\ }\textbf {\bibinfo {volume} {91}},\ \bibinfo {pages}
  {201102} (\bibinfo {year} {2003})}\BibitemShut {NoStop}%
\bibitem [{\citenamefont {{Sullivan}}\ \emph {et~al.}(2016)\citenamefont
  {{Sullivan}}, \citenamefont {{O'Connor}}, \citenamefont {{Zegers}},
  \citenamefont {{Grubb}},\ and\ \citenamefont {{Austin}}}]{sullivan-16}%
  \BibitemOpen
  \bibfield  {author} {\bibinfo {author} {\bibfnamefont {C.}~\bibnamefont
  {{Sullivan}}}, \bibinfo {author} {\bibfnamefont {E.}~\bibnamefont
  {{O'Connor}}}, \bibinfo {author} {\bibfnamefont {R.~G.~T.}\ \bibnamefont
  {{Zegers}}}, \bibinfo {author} {\bibfnamefont {T.}~\bibnamefont {{Grubb}}}, \
  and\ \bibinfo {author} {\bibfnamefont {S.~M.}\ \bibnamefont {{Austin}}},\
  }\href {\doibase 10.3847/0004-637X/816/1/44} {\bibfield  {journal} {\bibinfo
  {journal} {\apj}\ }\textbf {\bibinfo {volume} {816}},\ \bibinfo {eid} {44}
  (\bibinfo {year} {2016})},\ \Eprint {http://arxiv.org/abs/1508.07348}
  {arXiv:1508.07348 [astro-ph.HE]} \BibitemShut {NoStop}%
\bibitem [{\citenamefont {{Pascal}}\ \emph {et~al.}(2020)\citenamefont
  {{Pascal}}, \citenamefont {{Giraud}}, \citenamefont {{Fantina}},
  \citenamefont {{Gulminelli}}, \citenamefont {{Novak}}, \citenamefont
  {{Oertel}},\ and\ \citenamefont {{Raduta}}}]{pascal-20}%
  \BibitemOpen
  \bibfield  {author} {\bibinfo {author} {\bibfnamefont {A.}~\bibnamefont
  {{Pascal}}}, \bibinfo {author} {\bibfnamefont {S.}~\bibnamefont {{Giraud}}},
  \bibinfo {author} {\bibfnamefont {A.~F.}\ \bibnamefont {{Fantina}}}, \bibinfo
  {author} {\bibfnamefont {F.}~\bibnamefont {{Gulminelli}}}, \bibinfo {author}
  {\bibfnamefont {J.}~\bibnamefont {{Novak}}}, \bibinfo {author} {\bibfnamefont
  {M.}~\bibnamefont {{Oertel}}}, \ and\ \bibinfo {author} {\bibfnamefont
  {A.~R.}\ \bibnamefont {{Raduta}}},\ }\href {\doibase
  10.1103/PhysRevC.101.015803} {\bibfield  {journal} {\bibinfo  {journal}
  {\prc}\ }\textbf {\bibinfo {volume} {101}},\ \bibinfo {eid} {015803}
  (\bibinfo {year} {2020})},\ \Eprint {http://arxiv.org/abs/1906.05114}
  {arXiv:1906.05114 [astro-ph.HE]} \BibitemShut {NoStop}%
\bibitem [{\citenamefont {{Hanke}}\ \emph {et~al.}(2013)\citenamefont
  {{Hanke}}, \citenamefont {{M{\"u}ller}}, \citenamefont {{Wongwathanarat}},
  \citenamefont {{Marek}},\ and\ \citenamefont {{Janka}}}]{hanke-13}%
  \BibitemOpen
  \bibfield  {author} {\bibinfo {author} {\bibfnamefont {F.}~\bibnamefont
  {{Hanke}}}, \bibinfo {author} {\bibfnamefont {B.}~\bibnamefont
  {{M{\"u}ller}}}, \bibinfo {author} {\bibfnamefont {A.}~\bibnamefont
  {{Wongwathanarat}}}, \bibinfo {author} {\bibfnamefont {A.}~\bibnamefont
  {{Marek}}}, \ and\ \bibinfo {author} {\bibfnamefont {H.-T.}\ \bibnamefont
  {{Janka}}},\ }\href {\doibase 10.1088/0004-637X/770/1/66} {\bibfield
  {journal} {\bibinfo  {journal} {\apj}\ }\textbf {\bibinfo {volume} {770}},\
  \bibinfo {eid} {66} (\bibinfo {year} {2013})},\ \Eprint
  {http://arxiv.org/abs/1303.6269} {arXiv:1303.6269 [astro-ph.SR]} \BibitemShut
  {NoStop}%
\bibitem [{\citenamefont {Mirizzi}\ \emph {et~al.}(2016)\citenamefont
  {Mirizzi}, \citenamefont {Tamborra}, \citenamefont {Janka}, \citenamefont
  {Saviano}, \citenamefont {Scholberg}, \citenamefont {Bollig}, \citenamefont
  {H{\"u}depohl},\ and\ \citenamefont {Chakraborty}}]{Mirizzi:2015eza}%
  \BibitemOpen
  \bibfield  {author} {\bibinfo {author} {\bibfnamefont {A.}~\bibnamefont
  {Mirizzi}}, \bibinfo {author} {\bibfnamefont {I.}~\bibnamefont {Tamborra}},
  \bibinfo {author} {\bibfnamefont {H.-T.}\ \bibnamefont {Janka}}, \bibinfo
  {author} {\bibfnamefont {N.}~\bibnamefont {Saviano}}, \bibinfo {author}
  {\bibfnamefont {K.}~\bibnamefont {Scholberg}}, \bibinfo {author}
  {\bibfnamefont {R.}~\bibnamefont {Bollig}}, \bibinfo {author} {\bibfnamefont
  {L.}~\bibnamefont {H{\"u}depohl}}, \ and\ \bibinfo {author} {\bibfnamefont
  {S.}~\bibnamefont {Chakraborty}},\ }\href {\doibase
  10.1393/ncr/i2016-10120-8} {\bibfield  {journal} {\bibinfo  {journal} {Riv.
  Nuovo Cim.}\ }\textbf {\bibinfo {volume} {39}},\ \bibinfo {pages} {1}
  (\bibinfo {year} {2016})},\ \Eprint {http://arxiv.org/abs/1508.00785}
  {arXiv:1508.00785 [astro-ph.HE]} \BibitemShut {NoStop}%
\bibitem [{\citenamefont {Melson}\ \emph {et~al.}(2015)\citenamefont {Melson},
  \citenamefont {Janka}, \citenamefont {Bollig}, \citenamefont {Hanke},
  \citenamefont {Marek},\ and\ \citenamefont {Müller}}]{Melson:2015spa}%
  \BibitemOpen
  \bibfield  {author} {\bibinfo {author} {\bibfnamefont {T.}~\bibnamefont
  {Melson}}, \bibinfo {author} {\bibfnamefont {H.-T.}\ \bibnamefont {Janka}},
  \bibinfo {author} {\bibfnamefont {R.}~\bibnamefont {Bollig}}, \bibinfo
  {author} {\bibfnamefont {F.}~\bibnamefont {Hanke}}, \bibinfo {author}
  {\bibfnamefont {A.}~\bibnamefont {Marek}}, \ and\ \bibinfo {author}
  {\bibfnamefont {B.}~\bibnamefont {Müller}},\ }\href {\doibase
  10.1088/2041-8205/808/2/L42} {\bibfield  {journal} {\bibinfo  {journal}
  {Astrophys. J. Lett.}\ }\textbf {\bibinfo {volume} {808}},\ \bibinfo {pages}
  {L42} (\bibinfo {year} {2015})},\ \Eprint {http://arxiv.org/abs/1504.07631}
  {arXiv:1504.07631 [astro-ph.SR]} \BibitemShut {NoStop}%
\bibitem [{\citenamefont {{The HDF Group}}(2020)}]{hdf5}%
  \BibitemOpen
  \bibfield  {author} {\bibinfo {author} {\bibnamefont {{The HDF Group}}},\
  }\href@noop {} {\enquote {\bibinfo {title} {{Hierarchical Data Format,
  version 5}},}\ } (\bibinfo {year} {1997-2020}),\ \bibinfo {note}
  {http://www.hdfgroup.org/HDF5/}\BibitemShut {NoStop}%
\end{thebibliography}%

\end{document}